\documentclass[journal]{IEEEtran}
\usepackage{xcolor,soul,framed} 

\colorlet{shadecolor}{yellow}
\usepackage[pdftex]{graphicx}

\graphicspath{{../pdf/}{../jpeg/}}
\DeclareGraphicsExtensions{.pdf,.jpeg,.png}

\usepackage[cmex10]{amsmath}
\usepackage{array}
\usepackage{mdwmath}
\usepackage{mdwtab}
\usepackage{eqparbox}
\usepackage{url}
\usepackage{algorithm}
\usepackage{algorithmic}
\usepackage{cite}
\usepackage{booktabs} 
\usepackage{multirow}
\usepackage{multicol}
\usepackage{amsfonts}
\usepackage{float}
\usepackage{bbm}
\usepackage{balance}
\usepackage{xcolor}
\usepackage{amssymb}
\usepackage{pifont}
\newcommand{\xmark}{\ding{55}}%

\begin{document}
    \title{Audio-Visual Target Speaker Extraction with \\ Selective Auditory Attention}
    \author{Ruijie Tao, Xinyuan Qian,~\IEEEmembership{Senior Member,~IEEE}, Yidi Jiang, Junjie Li, Jiadong Wang and \\ Haizhou Li,~\IEEEmembership{Fellow,~IEEE}
    
    \thanks{This work is supported by FD-fAbrICS: Joint Lab for FD-SOI Always-on Intelligent \& Connected Systems (I2001E0053); National Natural Science Foundation of China (62271432); Shenzhen Science and Technology Program (ZDSYS20230626091302006); Shenzhen Science and Technology Research Fund, Fundamental Research Key Project (JCYJ20220818103001002). (\textit{Corresponding author: Jiadong Wang})}
    
    \thanks{Ruijie Tao, Yidi Jiang and Jiadong Wang are with the Department of Electrical and Computer Engineering, National University of Singapore, Singapore (e-mail: ruijie.tao@u.nus.edu, yidi\_jiang@u.nus.edu and jiadong.wang@u.nus.edu) (\textit{Corresponding author: Jiadong Wang}).}
    \thanks{Xinyuan Qian is with the Department of Computer Science and Technology, University of Science and Technology Beijing, Beijing, 100083, China. (email: qianxy@ustb.edu.cn)}
    \thanks{Junjie Li is with the Chinese University of Hong Kong (Shenzhen), 518172 China (e-mail: mrjunjieli@gmail.com).}
    \thanks{Haizhou Li is with the Guangdong Provincial Key Laboratory of Big Data Computing, the Chinese University of Hong Kong (Shenzhen), 518172 China; also with  Shenzhen Research Institute of Big data, Shenzhen, 518172 China; also with the Department of Electrical and Computer Engineering, National University of Singapore, 119077, Singapore and also with the University of Bremen, 28359 Germany. (e-mail: haizhouli@cuhk.edu.cn)}
}
    
\maketitle

\begin{abstract}
Audio-visual target speaker extraction (AV-TSE) aims to extract the specific person's speech from the audio mixture given auxiliary visual cues. Previous methods usually search for the target voice through speech-lip synchronization. However, this strategy mainly focuses on the existence of target speech, while ignoring the variations of the noise characteristics, i.e., interference speaker and the background noise. That may result in extracting noisy signals from the incorrect sound source in challenging acoustic situations. To this end, we propose a novel selective auditory attention mechanism, which can suppress interference speakers and non-speech signals to avoid incorrect speaker extraction. By estimating and utilizing the undesired noisy signal through this mechanism, we design an AV-TSE framework named Subtraction-and-ExtrAction network (SEANet) to suppress the noisy signals. We conduct abundant experiments by re-implementing three popular AV-TSE methods as the baselines and involving nine metrics for evaluation. The experimental results show that our proposed SEANet achieves state-of-the-art results and performs well for all five datasets. The code can be found in: \textcolor{magenta}{\url{https://github.com/TaoRuijie/SEANet.git}}

\end{abstract}

\section{Introduction}

\begin{figure}[!ht]
  \centering
  \includegraphics[width=\linewidth]{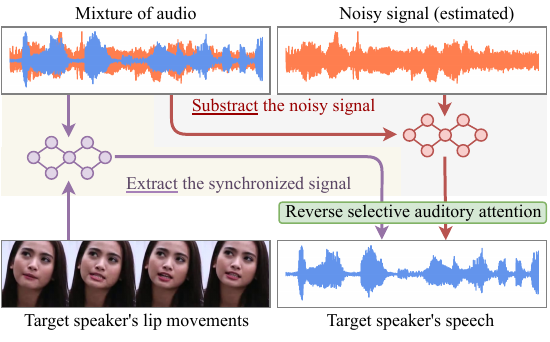}
  \caption{Typical approaches in AV-TSE focus on `Extraction', which searches for the target speaker's voice from the audio mixture to match the corresponding lip movements. However, the extraction results may involve noisy signals from incorrect sound sources. To alleviate these problems, we introduce the complementary `\textit{subtraction}' strategy. By analyzing this selective auditory attention, our proposed method utilizes the estimated noisy signal and excludes them during extraction.}
  \label{Cover}
\end{figure}

\IEEEPARstart{T}{he} human brain is capable of selectively listening to a specific voice and filtering out other sounds in situations where there are multiple talkers and non-speech noises. This phenomenon is known as the `selective auditory attention'~\cite{cherry1953some, golumbic2013visual, cutler2000look}. Inspired by that, the audio-visual target speaker extraction (AV-TSE) task has been studied to mimic this ability. In AV-TSE, neural networks can comprehend the lip movements of the target speaker and extract the corresponding clean speech~\cite{afouras2018conversation} from the heard mixture of audio. In practice, AV-TSE does not require prior knowledge of the speaker amount~\cite{pan2021muse, Lin2023sepformer} and relies on synchronized facial expressions to focus on the target speaker's voice only. As an essential front-end, AV-TSE can assist various downstream multi-modal tasks such as audio-visual speech recognition~\cite{wang2022predict, shi2022avhubert}, speaker diraization~\cite{cheng2024multi, tao2021someone} and recognition~\cite{tao2023self, cai2022incorporating}. It can also be applied to augmented reality (AR) glasses and robots to improve the quality of the heard voice~\cite{dianatfar2021review}.

The foundation of AV-TSE lies in the synchronization between lip movements and speech waveform~\cite{wu2019time, li23ja_interspeech}. In an audio mixture, the target speech inside represents the pronunciation of phonemes, while the lip movements of the target speaker correspond to the viseme signals~\cite{edelman1987neural, crosse2016eye}\footnote{Viseme: a basic visual speech unit links to the phoneme.}. By establishing a temporal correlation between visemes and the corresponding speech signal components, it becomes possible to associate and align these two different modal signals~\cite{bear2017phoneme}. Following this principle, an efficient AV-TSE solution is to encode visual lip frames into the embedding sequence, and feed it into the extractor module for obtaining the clean speech waveform. In particular, different architectures have been studied, such as convolutional neural networks (CNN)~\cite{afouras2018conversation, gao2021visualvoice, li2023an}, temporal convolutional networks (TCN)~\cite{luo2019conv, wu2019time, pan2021muse, pan2022selective}, cross-modal transformer~\cite{vaswani2017attention, Lin2023sepformer} and dual-path recurrent neural network (DPRNN)~\cite{luo2020dual, pan2022usev}.

To achieve robust extraction, AV-TSE model needs to search for the clean target speech (synchronized to lip movements) and remove the noisy signal (unsynchronized to lip movements). However, based on our previous study of speech-lip synchronization~\cite{tao2021someone}, a well-trained neural network can tolerate the noisy signal during synchronization detection. As depicted in the left part of Figure 1, most existing AV-TSE methods are designed according on the `\textit{extraction}' strategy only. The emphasis is to guarantee that the output contains the speech waveform that be time-aligned with lip movements~\cite{pan2021muse, Lin2023sepformer, afouras2018conversation}. Based on that, AV-TSE models may inadvertently involve noisy signals or even separate some voices from the wrong source~\cite{zhao2022target}. That inspires us to guide the model in perceiving noisy signals during extraction. Studies of negative prompting~\cite{miyake2023negative, li2024learning} can guide the generation by providing unwanted signals. However, the noisy signals are unknown in AV-TSE, and the speech-noisy correlation is a complex temporal correlation. Additionally, some studies in image denoising and speech recognition have revealed the significance of investigating noisy signals as the auxiliary~\cite{chen2018reverse, zheng2021interactive, hu2022interactive}. These motivate us to design a dedicated module that explicitly fits the AV-TSE task for exploring the speech-noise mutual exclusivity, i.e., selective auditory attention. We summarise that as the `\textit{subtraction}' strategy (as indicated by the right part of Figure~\ref{Cover}), which aims to estimate and understand the noisy signal in AV-TSE to improve training efficiency and system performance.

In this paper, we propose a reverse attention mechanism to exclude noisy information in AV-TSE to achieve this mutual exclusivity. It takes the target speech and the noisy signal for frame-level cross-attention. Suppose two audio frames from the respective signal achieve a high correlation score, these components should be penalized for incorrect extraction. This design can remove undesirable audio and avoid incorrect speaker extraction. Meanwhile, considering that noise includes interference speakers and background audio, which are all out-of-sync with the target speaker's lip movements, we design a parallel speech and noise learning (PSNL) block to estimate speech and noisy signals at the same time. Furthermore, to extend this idea to the multi-modal processing, we design a multi-modal temporal attention module with diverse attempts to explore the effect of audio-visual correlation. By leveraging these modules, the framework can perceive the noisy signal during the extraction process by combining the `\textit{extraction}' and `\textit{subtraction}' strategies.

The contributions in this paper can be listed as:
\begin{enumerate}
\item To avoid the presence of noisy information and incorrect speech in audio-visual target speaker extraction, we propose a specific reverse attention module to achieve the mutual exclusivity between the target speaker's speech and the noisy signal.
\item Based on that, we propose the Subtraction-and-ExtrAction network (\textit{SEANet}), a novel AV-TSE system with selective auditory attention to assist the extraction by excluding the noisy signal. SEANet contains the PSNL block to estimate the speech and noisy signal. We also design the SEANet+ framework with the multi-modal temporal attention to explore the effect of visual information.
\item We conduct the results on five datasets for in-domain and cross-domain evaluation. With solid comparison by re-implementing three baseline methods, SEANet achieves state-of-the-art results for all nine metrics.
\end{enumerate}

\section{Related work}
\subsection{Target speaker extraction}
In social activity scenarios, the auditory environment often comprises a complex mixture of several speakers' voices with reverberation and background noise~\cite{wang2018voicefilter, delcroix2018single}. Traditional speech separation (SS) aims to estimate the individual speech sources of each speaker from the mixture signal directly~\cite{wang2018supervised, luo2020dual, subakan2021attention}. It struggles with assigning separated speech to the correct speakers~\cite{yu2017permutation}. Inspired by human perception~\cite{xu2019time}, target speaker extraction (TSE) treats the person of interest as the target speaker to imitate selective auditory attention ability: In TSE, An auxiliary reference from the target speaker is provided to direct the target person. By leveraging this reference, the neural network can effectively extract the target speaker's voice from the audio mixture~\cite{xu2020spex, ge2020spex+, pan2021muse}. TSE does not require prior information of the total speaker number~\cite{xu2019time}. Meanwhile, benefits from the reference signal to supervise the extraction, TSE can prevent unreliable blind source separation. 

\subsection{Auxiliary references in target speaker extraction}

\begin{figure}[!tb]
  \centering
  \includegraphics[width=.95\linewidth]{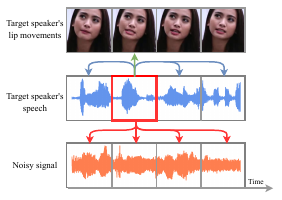}
  \caption{Available auxiliary references in AV-TSE. Green line: speech-lip synchronization (positive correlated); Blue line: voice consistency (positive correlated); Orange line: speech-noise exclusivity (negative correlated).}
  \label{Related_work}
\end{figure}

\subsubsection{Pre-enrolled speech} 
The widely-used auxiliary reference is the target speaker's pre-enrolled speech utterance~\cite{xu2020spex, ge2020spex+}. Each speaker possesses the distinctive voice characterizes, which can be modeled into a fixed dimensional vector, referred to as the speaker embedding~\cite{dehak2010front, snyder2016deep, wan2018generalized}. Previous works~\cite{wang2018voicefilter, xu2019time, mu2024self} have employed the speaker encoder to derive speaker embedding from the reference speech, subsequently serving as attractors in the extraction process~\cite{zhang2020x, ge2022spex}. Inspired by that, we propose to leverage the inherent consistency of the target speaker's voice for AV-TSE. As illustrated by the blue lines in Figure~\ref{Related_work}, we expect that each frame of the extracted speech exhibits a high degree of correlation with the others due to the consistent voiceprint.

\subsubsection{Speech-lip synchronization} 
The facial information, particularly the lip movements of the target speaker, can be used as the auxiliary reference in extraction through speech-lip synchronicity~\cite{afouras2018conversation}. Following this idea, several AV-TSE algorithms have been proposed, such as MuSE for self-enrollment~\cite{pan2021muse}, USEV for sparse overlap scenario~\cite{pan2022usev} and TDSE for time-domain modeling~\cite{wu2019time}. These methods use a visual encoder to learn lip embedding sequences of the target speaker, which are then temporally aligned with speech frames. As shown by the green lines in Figure~\ref{Related_work}, during extraction, the interaction occurs between the concurrent speech frames and the visual lip frames~\cite{li23ja_interspeech, sato2021multimodal, li2023an}.

\subsubsection{Speech-noise exclusivity}
Noise always comes with speech, this correlation (speech-noise exclusivity) can provide complementary guidance for extraction. For example, in computer vision, extracting noise from images can contribute to the denoising process~\cite{lin2021noise2grad}; In speech recognition and enhancement, the consideration of noise can improve the system through a lightweight interaction layer~\cite{zheng2021interactive, hu2022interactive}. These findings demonstrate the importance of involving noise exploration. However, as illustrated by the red lines in Figure~\ref{Related_work}, this inter-frame mutual exclusivity between the target speaker's speech and the noisy signal (refer as the `selective auditory attention' mechanism) has not been explored in AV-TSE. 

\subsubsection{Utilization of the negative singal}
Utilizing the negative signal is verified to be efficient in multiple tasks. For instance, in text-to-image generation, the negative prompting solution~\cite{miyake2023negative, li2024learning} can guide the generation process by providing the prompt to describe the unwanted signal. Meanwhile, previous studies~\cite{chen2018reverse, fan2020pranet} in object detection and polyp segmentation~\footnote{the process of delineating the boundary of a polyp from the colon wall} studied the concept of reverse attention, which eliminates the core regions to make the model focus on local details. However, compared with these non-temporal relationship, the speech-interference correlation is a more complex temporal relationship with the unknown interference presentation. Inspired by that and modified based on the standard self-attention mechanism, we propose the reverse attention module. It penalizes segments with high speech-noise and speech-vision consistency since they should be mutually exclusive.

\vspace{4mm}

\section{SEANet}

\subsection{Task formulation and data generation}

\begin{figure}[!ht]
  \centering
  \includegraphics[width=.98\linewidth]{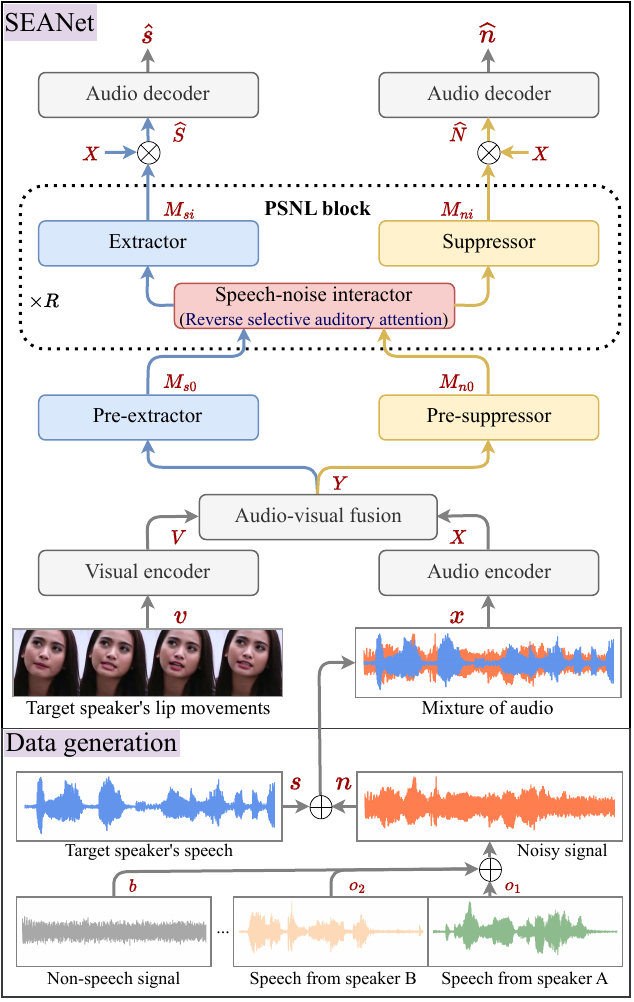}
  \caption{The bottom panel is the data generation process for AV-TSE, the upper panel is our proposed SEANet. It extracts the clean speech of the target speaker from the mixture of audio. Specifically, SEANet contains $R$ repeated PSNL blocks to learn the selective auditory attention between the estimated clean speech and noisy signal, $M_{si}$ and $M_{ni}$ are the output speech and noise embeddings from the $i^{th}$ block, respectively. $\oplus$ represents entrywise sum (to mix up audios) and $\otimes$ represents matmul product. Note that Pre-extractor and Pre-suppressor share the same model architecture with Extractor and Suppressor, respectively.}
  \label{SEANet}
\end{figure}

We define $x$ as a mixture of audio in the time domain, which contains the target speaker's voice $s$, the interference speaker's voice $o_i$ and the background non-speech signal $b$:

\begin{equation}
    x = s + \sum_{i=1}^N{o_i} + b
\end{equation}
where $N$ is the number of interference speakers that $\geq 1$. As shown in Figure~\ref{SEANet}, in data generation process,
we sum the speech from other speakers and the non-speech signal to generate the noisy signal: $n = \sum_{i=1}^N{o_i} + b$. Then $x$ can be obtained by summing $n$ and $s$. Given the visual frames of the target speaker, which are denoted as $v$, AV-TSE aims to remove all noisy signal $n$ from $x$ and obtain $\hat{s}$ to approximate $s$. $v$ and $s$ should be synchronized.
In our proposed SEANet, $v$ and $x$ are fed as the inputs which are encoded into the visual and audio embeddings to be fused and segmented into the multimodal embedding.
In order to obtain the embedding of the target speaker's speech and the noisy signal for the subsequent interaction, the pre-extractor and  pre-suppressor modules are designed. Then several parallel speech and noise learning (PSNL) blocks are serially connected to learn the selective auditory attention. Each PSNL block contains the reverse attention module to penalize the incorrect extraction. Finally, the embedding of the target speaker's speech and the noisy signal are aggregated and decoded into the estimated target speaker's speech $\hat{s}$ and the estimated noisy signal $\hat{n}$. Note that only $\hat{s}$ is required during evaluation. 

\subsection{Audio and visual encoder} 
For audio input, the mixture of input audio $x$ is the 1-dimensional (1D) signal in the time domain with the length of $T$. The audio encoder is viewed as a frequency analyzer to output the 2D frame-based embedding sequence $X \in R^{D_a \times L}$. This encoder consists of a 1D-CNN and a rectified linear activation (ReLU). $D_a$ and $L$ is the audio embedding dimension and the number of audio frames, respectively. The duration of each audio frame is decided by 1D CNN's stride size. 

For visual inputs, a CNN-based visual encoder is utilized to discern lip movements from static face frames, with a focus on the specific mouth shape. The target speaker's face frames 
$v$ are processed, where a central crop is applied to isolate and efficiently represent the lip region.
The visual frame representations are generated using a 3D-CNN layer followed by a ResNet18 block, as detailed in~\cite{afouras2018deep}. Subsequently, the Video Temporal Convolutional block (V-TCN) is engaged to extract temporal dynamics of lip movements and to identify the corresponding viseme.
This V-TCN comprises a sequence of five residual connections, each integrating a Rectified Linear Unit (ReLU), batch normalization (BN), and a depth-wise separable convolutional layer (DS-Conv1D), as outlined in~\cite{wu2019time}.
The output embedding is then up-sampled to $V \in R ^ {D_v \times L}$, aligning with the temporal resolution of the audio embedding $X$ ($D_v$ denotes the embedding dimension).

\subsection{Audio-visual fusion}
Previous research has confirmed that temporally aligning and concatenating audio and visual features is an effective approach for feature extraction~\cite{wu2019time, pan2021muse}. Based on that, the audio embedding $X$ is passed through a group normalization (GN) and a 1D-CNN layer. Subsequently, this embedding is concatenated with visual embedding $V$ along the temporal dimension, which is then further refined by an additional 1D-CNN layer. The dimension of the integrated audio-visual embedding is $R ^ {D \times L}$ where $D$ represents the embedding dimension, and $L$ is the number of audio frames, respectively.

To manage long feature sequences of variable lengths, the segmentation operation effectively divides the audio-visual embedding into multiple, overlapping segments along the temporal axis, as described in~\cite{luo2020dual}.  The segmentation utilizes a window size of $K$ and a hop size of $K/2$, yielding $P$ segments, which are concatenated to create a 3D tensor $Y \in R ^ {D \times K \times P}$. 
After we obtain the embedding of the clean target speech $M_{si}$ (which will be introduced later), an aggregation step, acting as the inverse of segmentation, reconstitutes the 3D tensor into a 2D tensor. 
It undergoes a parametric rectified linear unit (pReLU) activation and passes through a 1D-convolutional (1D-Conv) layer to derive the estimated speech embedding $\hat{S}$.
A parallel process is employed to attain the estimated noisy embedding $\hat{N}$. Noted that the segmentation and aggregation processes are not explicitly illustrated in Figure~\ref{SEANet}.

\subsection{Pre-extractor and pre-suppressor} 
To achieve the subsequent speech-noise interaction process, we initially estimate the embeddings of the target speaker's speech and the accompanying noise. 
As depicted in Figure~\ref{SEANet}, the audio-visual embedding $Y$ is fed into the pre-extractor and pre-suppressor to obtain the  initial embedding $M_{s0}$ for the target speech and  $M_{n0}$ for the noisy signal, respectively. 
These pre-processing modules are adaptations of DPRNN~\cite{luo2020dual}. Note that the extractor and suppressor (which are elaborated in the subsequent PSNL section) share identical model architectures with their pre-processing counterparts. More specifically, the extractor in Figure 4 has the same architecture as the pre-extractor, and the suppressor in Figure 4 has the same architecture as the pre-suppressor.

Pre-extractor contains an intra-chuck module followed by an inter-chuck module. 
In the intra-chunk module, a Bidirectional Long Short-Term Memory (BLSTM) layer is applied to integrate the knowledge within each chunk. The length of each chunk, denoted as $K$ , is considered as the sequence length for the BLSTM.
This BLSTM is followed by a linear layer with GN. 
The inter-chuck module has a similar network structure. The difference is that the sequence length in BLSTM is the number of chunks $P$. This enables the knowledge integration across the different chunks. 

Pre-suppressor is designed to estimate the embedding of the noisy signal. It has the same model structure as the extractor to search for the out-of-sync audio signal with the visual frames. This suppressor works in parallel with the extractor.

\begin{figure}[!htb]
  \centering
  \includegraphics[width=\linewidth]{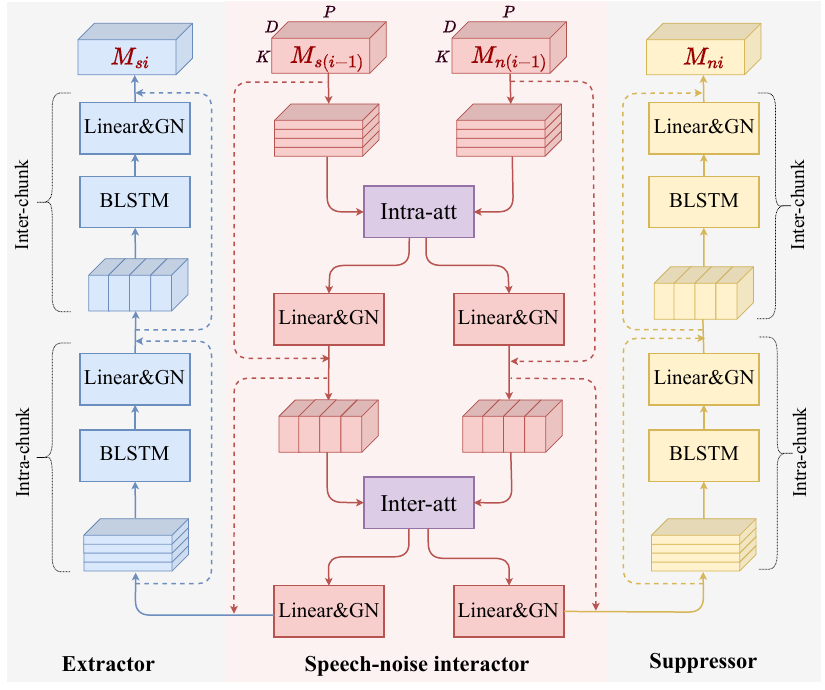}
  \vspace{1mm}
  \caption{The $i^{th}$ parallel speech and noise learning (PSNL) block. Intra-att and Inter-att blocks are used to learn the interaction between the target speaker's speech and the noisy signal. $M_{si}$ and $M_{ni}$ are the output embeddings of the speech and the noise, respectively. Dash lines denote residual connections. Pre-extractor/suppressor in Fig~\ref{SEANet} has the same structure as the extractor/suppressor, while the input is $Y$, and the output is $M_{s0}$ or $M_{n0}$. }
  \label{PSNL}
\end{figure}

\subsection{Parallel speech and noise learning (PSNL)}

After that, parallel speech and noise learning (PSNL) block is used to learn the selective auditory attention. $M_{s0}$ and $M_{n0}$ are fed into $R$ PSNL blocks. These blocks have non-shared parameter weights and are serially connected. In the $i^{th}$ PSNL block, the inputs are the speech embedding $M_{s(i-1)}$ and the noise embedding $M_{n(i-1)}$ from the $(i-1)^{th}$ PSNL block ($i \in [1,R]$). First, the intra-chunk attention (intra-att) module is used to learn the speech-noise mutual exclusivity within each chunk, which can avoid incorrect extraction in local frames. The output embeddings go through the linear layers with GN operation. Then, with a similar structure, an inter-att module is applied to compare the speech and noisy signals from different chunks to learn the global correctness. Here, the length of each chunk $K$ and the number of chunks $P$ are viewed as the sequence length in the intra-att and inter-att module, respectively. As shown in Figure~\ref{PSNL}, after the speech-noise interactor, an extractor and a suppressor are used to optimize the embedding of the target speaker's speech and the noisy signal. This extractor (suppressor) have the same model architecture to the pre-extractor (pre-suppressor).

\subsection{Reverse attention module} 
\begin{figure}[!tb]
  \centering
  \includegraphics[width=\linewidth]{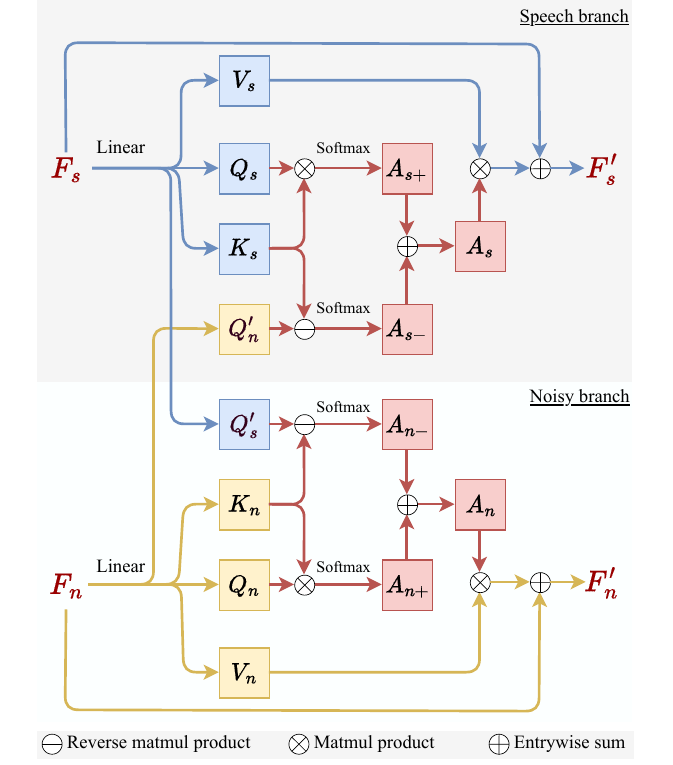}
  \caption{Illustration of the intra-chunk attention module (intra-att) or the inter-chunk attention module (inter-att). $F_s$ and $F_n$ represent the input speech and noisy embedding, respectively. $F^{\prime}_{s}$  and $F^{\prime}_{n}$ denote their respective processing outputs. This module contains the self-attention score $A_{s+}$ to learn the speech-lip relationship and the reverse attention score $A_{s-}$ to comprehend the speech-noise mutual exclusivity. }
  \vspace{-4mm}
  \label{Attention_module}
\end{figure}

Figure~\ref{Attention_module} explains the details of the intra-att and inter-att module. They have the similar structure, and each integrating a self-attention component and a reverse attention component. 
Specifically, the self-attention mechanism assigns higher weights to audio frames that exhibit a strong correlation to enhance the voice consistency. 
Conversely, the reverse attention mechanism conducts a frame-level cross-attention analysis between the target speech signal and the accompanying noise.
If two audio frames from the respective signal achieve a high correlation score, that represents the extracted speech segment is similar to the estimated noisy signal. In other words, the extracted speech may contain noise and should be penalized. 

We designate the inputs to the intra-chunk attention module as $F_s$, representing the 3D speech embedding, and $F_n$, representing the noise embedding.
Four linear layers are used for the speech branch to obtain $V_s$, $Q_s$, $K_s$ and $Q^{\prime}_s$. Here, $Q_s$ serves as the query vector in self-attention to ensure voice consistency. $Q^{\prime}_s$ operates as the query vector in reverse attention for speech-noise exclusivity. $K_s$ and $V_s$ are the key and value of speech, respectively. With a similar approach, $V_n$, $Q_n$, $K_n$ and $Q^{\prime}_n$ can be obtained from the noise embedding $F_n$.

For self-attention, the extracted speech segments should exhibit consistent voice characteristics across the temporal dimension. To this end, we multiply the query and key from speech to get the attention matrix $A_{s+}$. This matrix awards the extracted segment if it shares a consistent voice representation in an utterance. 

For reverse attention, the extracted speech should exclude the interference signal with low correlation: After multiplying the speech key and the noisy query, the output attention matrix goes through a reverse operation for the opposite attention matrix $A_{s-}$. Based on that, the high-corrected parts will be transferred into a low negative score to represent the incorrect extraction. The subsequent softmax operation $softmax(\cdot)$ can ensure that the final score matrix is in a suitable range. The final score matrix $A_{s}$ is defined as:
\begin{equation}
    A_{s} = \frac{1}{2} (\underbrace{softmax{(\frac{Q_s \cdot {K_s}^T}{\sqrt{D}}}}_{A_{s+}}) + \underbrace{softmax{(-\frac{Q^{\prime}_n \cdot {K_s}^T}{\sqrt{D}})}}_{A_{s-}})
    \label{Equation}
\end{equation}
which is the average of the self-attention score $A_{s+}$ and reverse attention score $A_{s-}$ to judge the quality of each extracted speech segment. It is then multiplied with the value of speech, followed by the residual connection to get the output embedding $F^{\prime}_{s}$. 

\begin{equation}
    F^{\prime}_{s} = A_{s} \cdot V_{s} + F_{s}
\end{equation}

As described in Figure~\ref{Attention_module}, the noisy branch has the symmetrical structure to obtain the $F^{\prime}_{n}$. Meanwhile, intra-att and inter-att share the similar structure, the difference is that the size of score matrix is $K \times K$ in intra-att and $P \times P$ in inter-att, respectively.

\subsection{Audio decoder and constraints design}

The final output embeddings are represented as $M_{sR}$ and $M_{nR}$. They are element-wise multiplied with $X$, respectively, followed by two audio decoder (shared weights) to obtain the estimated temporal target speech $\hat{s}_{R} \in R^{1 \times T}$ and noisy signal $\hat{n}_{R} \in R^{1 \times T}$. This decoder comprises a linear layer and an overlap-and-add (OLA) operation from signal processing.

We also multiply $M_{si}$ (and $M_{ni}$) with $X$ when $i \in [0, R-1]$, followed by the same audio decoder to obtain the speech signal $\hat{s}_i$ (and noisy signal $\hat{n}_i$) in each PSNL block. They will be involved in training loss to ensure that $M_{si}$ and $M_{ni}$ are direct to speech and noisy information, respectively, to guarantee meaningful speech-noise interaction. We use $\hat{s}_R$ (estimated target speaker's speech from the last PSNL block) to compute the metrics during evaluation.

\subsection{Loss function}
The scale-invariant signal-to-distortion ratio (SI-SDR) loss~\cite{le2019sdr} can measure the reconstruction error between the estimated audio $\hat{p}$ and the ground-truth audio $p$. We denote it by $l(\cdot, \cdot)$:

\begin{equation}
        l(\hat{p}, p) = - 10 \log_{10} ( \frac{||\frac{<\hat{p},p>p}{||p||^2}||^2}{||\hat{p} - \frac{<\hat{p},p>p}{||p||^2}||^2})
\end{equation}

Our training loss includes a main loss and an auxiliary loss. The main loss is the SI-SDR loss between the estimated target speaker's speech from the last PSNL block $\hat{s}_{R}$ and the ground-truth speech $s$:

\begin{equation}
        {\mathcal{L}_{main}} = l(\hat{s}_{R}, s)
\end{equation}

The auxiliary loss considers the outputs from the rest of the extractors and suppressors. It controls the inputs of interactor for training constraints:

\begin{equation}
        {\mathcal{L}_{aux}} = \sum_{i=0}^{R-1}{l(\hat{s}_i, s)} + \sum_{i=0}^R{l(\hat{n}_i, n})
\end{equation}

The final training loss is the weighted addition of these two losses. We also carry out experiments to study the robustness of this hyper-parameter $\beta$.

\begin{equation}
        \mathcal{L} = \mathcal{L}_{main} + \beta\cdot\mathcal{L}_{aux}
\end{equation}

\section{Exploration of multi-modal interaction}

\begin{figure}[!tb]
  \centering
  \includegraphics[width=\linewidth]{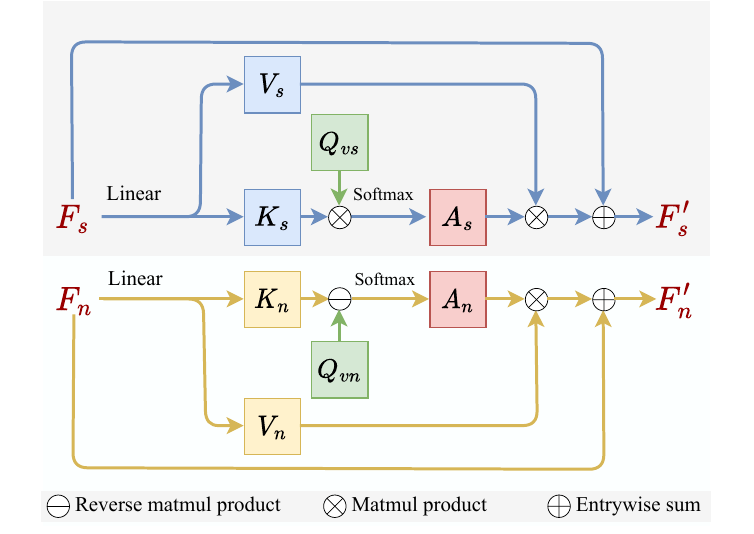}
  \caption{Multi-modal temporal attention block, which learns the interaction between the lip movements and the audio signal. F-SEANet system applies this module for audio-visual fusion, P-SEANet system inserts this module into the PSNL block after speech-noise interactor.}
  \vspace{-4mm}
  \label{Multimodal}
\end{figure}

Compared with the audio-only TSE, AV-TSE has the advantages of leveraging the audio-visual correlation. Motivated by that, we adapt our design and propose different multi-modal interaction methods to boost SEANet. This utilizes lip movements to assist the target speech extraction and the interference signals suppression. Three different systems are studied: F-SEANet applies multi-modal temporal attention for the audio-visual fusion step, P-SEANet inserts the multi-modal temporal attention into the PSNE block, after the speech-noise interactor; A-SEANet adds the additional contrastive loss during training.

\subsection{Fusion-based multi-modal interaction}

The audio-visual fusion block in SEANet concentrates the obtained signal modality signal along the temporal domain. As shown in Fig~\ref{Multimodal}, we design a multi-modal temporal attention to further explore our proposed reverse attention mechanism to explore more efficient audio-visual fusion.

Considering that the audio signal contains the target speech and the interference signal, in our multi-modal temporal attention, the visual signal interacts with the audio signal based on two branches: In both branches, the visual signal provides the query ($Q_{vs}$ and $Q_{vn}$), and the audio signal provides the key and value ($K_s$, $K_n$, $V_s$ and $V_n$). Here $F_{s} == F_{n}$, the difference is that one branch learns the positive audio-visual correlation to obtain the speech signal, while the other studies the negative audio-visual correlation and outputs the noisy signal. After that, these two signals ($F'_{s}$ and $F'_{n}$) are combined to get a robust audio representation. We name this system as F-SEANet.

\subsection{PSNL-based multi-modal interaction}

Then, we study the effect of conducting the multi-modal interaction into the PSNL block and name this system P-SEANet. First, to adjust the visual signal for multi-modal alignment, the output from the video encoder is segmented into several chunks to provide the visual query. After that, the multi-modal temporal attention module is inserted in each PSNL module, after each speech-noise interactor. This module contains architecture similar to the aforementioned one in the fusion block. The difference is that the inputs are the speech feature $F_{s}$ and the noisy feature $F_{n}$, and the outputs are the attentive speech and noisy feature ($F'_{s}$ and $F'_{n}$) without fusion.

\subsection{Algorithm-based multi-modal interaction}

Finally, we design the method to explore the multi-modal interaction from the algorithm aspect. Since the learned target speech has a positive correlation with the lip movements and the interference signal has a negative correlation, we design a contrastive learning based algorithm to involve the additional loss function and name it A-SEANet.

In A-SEANet, after each PSNL block, we first compute the cosine similarity between the speech and visual features along the time domain. Based on synchronization, these scores should be as high as possible; by contrast, the cosine similarity between the noisy and visual features ought to be as low as possible. These two similarity scores are subtracted as the additional training loss to update the system. We hope this can guide the framework in learning multi-modal temporal interaction.

\section{Experimental setting}

\subsection{Training set}

\subsubsection{LRS2} 
The LRS2 audio-visual speech recognition dataset~\cite{afouras2018deep} is derived from BBC television. Each video clip contains one person's synchronized audio-visual signal. To simulate one mixed audio, we randomly pick one sample's speech waveform and visual frames as the target speaker's clean speech and lip movements, respectively. The audio from one other sample is selected as the interference speech. This interference speech is mixed with the target speech with a random signal-to-noise (SNR) ratio between -10 dB and 10 dB. Similar to the previous studies~\cite{pan2021muse, Lin2023sepformer}, we generate 20,000, 5,000 and 3,000 mixed audios for training, validation and testing, respectively, with no overlapped speaker. We conduct the experiments with multiple people and the non-speech signal to simulate more diverse real-world scenarios in the quality studies.

\subsubsection{VoxCeleb2} 
VoxCeleb2~\cite{Voxceleb2} is an audio-visual speaker recognition dataset collected from YouTube interviews, which represents more challenging scenarios such as the unclear side face and the strong noise. The data generation pipeline is the same as in processing LRS2. 

For in-domain evaluation, our system is trained on VoxCeleb2 or LRS2 and then evaluated on the corresponding test set. Note that in all our experiments, the audio mixtures are simulated based on two speech signals, except for the experiment described in Table~\ref{tab:aug}.

\subsection{Evaluation Set}
For cross-domain evaluation, we train the system on VoxCeleb2 and test it on LRS3~\cite{afouras2018lrs3}, Grid~\cite{alghamdi2018corpus} and TCD-TIMIT~\cite{harte2015tcd}. During the evaluation, we randomly generate 3,000 speech mixtures in each evaluation dataset with variable lengths with the SNR between -10 dB to 10 dB. The details of these sets are listed as follows:

\subsubsection{LRS3}
LRS3 is the audio-visual speech dataset collected from the TED videos. As a large-scale speech recognition dataset, LRS3 contains videos with precise lip movements and clear speech. Furthermore, the audio and visual modalities are strictly synchronized. LRS3 can be used for indoor scenario evaluation of videos in the wild.

\subsubsection{Grid}
Grid dataset contains the videos from a controlled studio environment. As the multi-talker speech perception dataset for computational-behavioral studies, it has high-quality audio and video recordings. In each video, one speaker sits in front of the camera and reads the given sentence with a fixed head orientation.

\subsubsection{TCD-TIMIT}
TCD-TIMIT dataset also consists of the videos collected from the studio condition for speech recognition. In each video, two cameras with different angles are used to collect data, and the speaker is asked to read the given phonetically rich sentence. Both TCD-TIMIT and Gird can be employed for high-quality studio video evaluation.

\subsection{Implementation details}
During training, the initial learning rate is $10^{-3}$ with a decreasing interval of $3\%$ after every three training epochs. The maximum training epoch is set to 150, and the system is validated every three epochs. Videos are cut into segments with fixed duration during training (2 seconds in LRS2 and 4 seconds in VoxCeleb2). During the evaluation, the entire videos with the variable lengths are applied. 

The visual frame rate and audio sampling rate of all videos are 25fps and 16kHz, respectively. The visual inputs are the middle part of all face frames with the size of $112\times112$. That operation can improve training efficiency while keeping lip areas. Following all previous works~\cite{afouras2018conversation, wu2019time, pan2021muse, Lin2023sepformer}, the visual encoder is pre-trained by lip reading \footnote{\url{https://github.com/smeetrs/deep_avsr}} and it is frozen in all experiments. $K, D_a, D_v, D, R, \beta$ are set as 100, 256, 256, 64, 5 and 0.1, respectively. Our SEANet contains 8.70M parameters.

\subsection{Baseline methods}

To verify the robustness of our SEANet, we involve and re-implement three popular audio-visual speaker extraction systems as the baselines for comparison: AV-DPRNN, MuSE and AV-Sepformer. We build these network architectures and train them with the same setting that used for SEANet to achieve fair comparison.

\subsubsection{AV-DPRNN}
\begin{figure}[!htb]
  \centering
  \includegraphics[width=\linewidth]{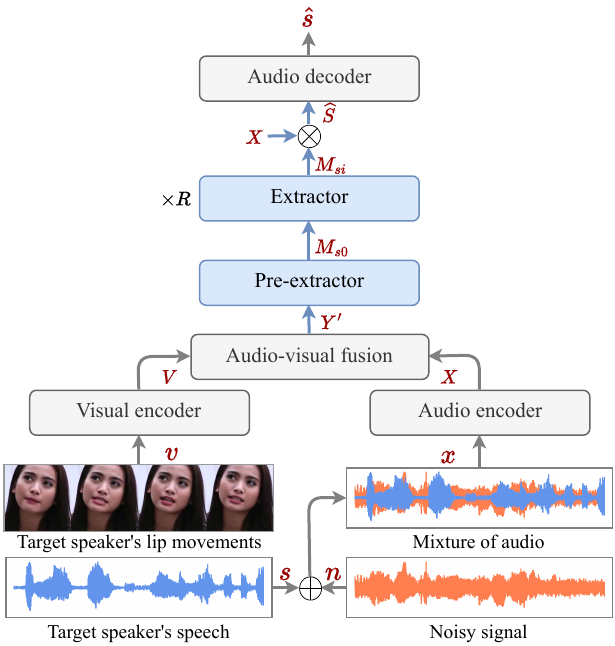}
  \caption{The AV-DPRNN framework is one of the baseline methods we re-implement. Compared with SEANet, it only contains extractor blocks to focus on the target speaker's speech.}
  \label{AV_DPRNN}
\end{figure}

AV-DPRNN is the audio-visual dual-path RNN-based TSE system. Compared to SEANet in Figure~\ref{SEANet}, the difference is that there is no suppressor, pre-suppressor and speech-noise interactor in the AV-DPRNN. As shown in Figure~\ref{AV_DPRNN}, it predicts the target speech only so the performance gap can reflect the effect of reverse attention. The loss function considers the output of all $(R+1)$ extractors. That setting is equivalent to the architecture of SEANet. AV-DPRNN has 4.12M parameters.

\subsubsection{MuSE}
MuSE (MUlti-modal Speaker Extraction network)~\cite{pan2021muse} is the temporal convolutional network-based system that be modified based on AV-ConvTasNet~\cite{wu2019time}. Considering that the voiceprint of the target speaker can also provide the reference for extraction, MuSE uses a self-enrollment strategy to learn the speaker embedding during the extraction process. The additional classification loss is applied during training to guide the system in understanding the speaker representation. It is trained on VoxCeleb2 for the original paper, and we re-implement the structure and conduct it on VoxCeleb2 and LRS2. Compared with SEANet, MuSE involves the additional speaker embedding information for the target speaker to guide the extraction. It contains 15.01M parameters, much larger than that in SEANet.

\subsubsection{AV-Sepformer}
AV-Sepformer~\cite{Lin2023sepformer} (AV-Sep) is the transformer-based multimodal extraction framework motivated by the transformer-based audio-only speech separation system. It introduces the audio-visual interaction to learn the relationship between the audio and visual modality. AV-Sepformer also contains a dual-path structure to understand inter-chunk and intra-chunk knowledge. Also, it was only studied on VoxCeleb2 for the original paper, and we extended the experiments for the LRS2 dataset. Considering that AV-Sepformer has a very large model architecture, for relatively fair comparison and GPU memory consideration, our study slightly reduced the number of Transformer layers in AV-Sepformer from 8 to 6. However, it still contains 21.73M parameters, which is around twice than that in SEANet.

\subsection{Metrics}

Following the previous works, we use scale-invariant signal-to-distortion ratio (\underline{SI-SDR})~\cite{le2019sdr} and signal-to-distortion ratio (\underline{SDR}) as the main metrics to evaluate the quality of the extracted speech. Their unit is dB. Meanwhile, we introduce several additional metrics to comprehensively judge the quality of the extraction: Perceptual evaluation of speech quality (\underline{PESQ})~\cite{rix2001perceptual} and short-term objective intelligibility (\underline{STOI})~\cite{taal2010short} are introduced to precept the extracted speech. Furthermore, we also report these metrics with the relative improvement (denoted as the suffix `$i$', i.e., \underline{SI-SDRi}, \underline{SDRi}, \underline{PESQi} and \underline{STOIi}). For instance, denote ground-truth speech as $s$, mixed audio as $x$, output speech as $\hat{s}$, there is $SDRi = SDR(\hat{s}, s) - SDR(x, s)$. Higher is better for these 8 metrics. Moreover, we also compute the word-error-rate (\underline{WER}) by feeding the extracted speech into the Whisper model (large version), which is the automatic speech recognition model~\cite{radford2023robust} that pre-trained on 680k hours of  diverse audio datasets. It can represent the readability and the informativeness of the extracted speech. Lower is better for WER.

\section{Results and analysis}
This section reports the results with five subsections: Section~\ref{res-1} compares SEANet with the baseline methods and other previous methods for in-domain evaluation. Both the results on LRS2 and VoxCeleb2 dataset have been provided. Section~\ref{res-2} reports the cross-domain evaluation result for SEANet and the baseline methods. The evaluation sets are LRS3, Grid and TCD-TIMIT datasets. In Section~\ref{res-3}, we conduct the ablation study to verify the efficiency of the architecture of SEANet, especially for reverse attention. Section~\ref{res-4} employs the quality study to prove the robustness of SEANet with different evaluation conditions and settings. Finally, section~\ref{res-5} provides some visualization results.

\subsection{\textbf{In-domain evaluation: LRS2 and VoxCeleb2}}
\label{res-1}

In Table~\ref{LRS2} and \ref{Vox2}, we conduct the experiments of SEANet and re-implement three pre-mentioned baselines with the same training setting for strictly fair comparison. For Table~\ref{others}, we provide the performance of other previous methods based on the results in their published paper and compare them with SEANet. Table I, II and III are used together to achieve a comprehensive and fair comparison with all previous solutions.

\subsubsection{\textbf{Comparison with baseline methods on the LRS2}}

Table~\ref{LRS2} contains the results of SEANet and baseline methods for all metrics. All four methods are trained and evaluated on LRS2 with the same setting by our effects. SEANet performs the 13.08 dB SI-SDR and 13.67 dB SDR, which is better than other methods, for instance, 12.60 dB SI-SDR from AV-Sep. SEANet also achieves the best PESQ and STOI (2.25 and 0.92). On the other hand, SEANet achieves better relative improvement on these four metrics. Furthermore, SEANet obtains a lower WER than the other three baseline methods, representing stronger readability. These results prove that our proposed SEANet can perform well for the AV-TSE task with a relatively lightweight model architecture and proper amount of training samples (20k).

\begin{table*}[!ht]
    \centering
        \caption{Comparison of SEANet and baseline methods for nine evaluation metrics, we train and evaluate them on the LRS2 dataset with the same setting. All results come from our own re-implement for fair comparison. Lower is better for WER and higher is better for the rest metrics.}
    \begin{tabular}{ccccccccccc}
        \hline
        \multirow{2}{*}{Method} & \multirow{2}{*}{\# Param} & \multicolumn{9}{c}{Metrics}\\
        &  & SI-SDR & SDR & PESQ & STOI & SI-SDRi & SDRi & PESQi & STOIi & WER (\%) \\
        \hline
         AV-DPRNN     & 4.12M & 10.24 & 10.81 & 1.88	& 0.88 & 10.29 & 10.65 & 0.70 & 0.22 & 14.81 \\
         MuSE~\cite{pan2021muse}         & 15.01M & 10.97 & 11.57 & 1.98 & 0.90 & 11.01 & 11.41 & 0.80 & 0.24 & 13.56 \\
         AV-Sep~\cite{Lin2023sepformer}       & 21.73M & 12.60 & 13.13 & 2.20 & 0.91 & 12.64 & 12.97 & 1.02 & 0.26 & 12.82 \\
         \textbf{SEANet}       & 8.70M & \textbf{13.08} & \textbf{13.67} & \textbf{2.25} & \textbf{0.92} & \textbf{13.05} & \textbf{13.52} & \textbf{1.07} & \textbf{0.29} & \textbf{12.26} \\
         \hline
    \end{tabular}
    \label{LRS2}
\end{table*}

\subsubsection{\textbf{Comparison with baseline methods on the VoxCeleb2}}

Table~\ref{Vox2} reports the performance of SEANet and baseline methods on the VoxCeleb2 dataset. Our SEANet performs the best SI-SDR and SDR on the VoxCeleb2, for instance, 12.88 dB SI-SDR from SEANet vs 10.84 SI-SDR from AV-Sep. Also, SEANet achieves the best PESQ and STOI with 2.34 and 0.89, respectively. Noted that the PESQ in VoxCeleb2 is slightly better than that in LRS2 because the videos in VoxCeleb2 have the longer duration. In summary, SEANet achieves an obvious improvement over the previous methods for all eight metrics.

\begin{table*}[!ht]
    \centering
        \caption{Comparison of SEANet and baseline methods for eight evaluation metrics, we train and evaluate them on the VoxCeleb2 dataset with the same setting. All results come from our own re-implement for fair comparison. Higher is better for all these metrics.}
        
    \begin{tabular}{cccccccccc}
        \hline
        \multirow{2}{*}{Method} & \multirow{2}{*}{\# Param} & \multicolumn{8}{c}{Metrics}\\
        &  & SI-SDR & SDR & PESQ & STOI & SI-SDRi & SDRi & PESQi & STOIi \\
        \hline
         AV-DPRNN     & 4.12M & 11.13 & 11.55 & 2.09 & 0.86 & 11.02 & 11.35 & 0.84 & 0.23  \\
         MuSE~\cite{pan2021muse}         & 15.01M & 10.89 & 11.34 & 2.11 & 0.86 & 10.78 & 10.51 & 0.86 & 0.23  \\
         AV-Sep~\cite{Lin2023sepformer}       & 21.73M & 12.07 & 12.49 & 2.25 & 0.88 & 11.96 & 12.29 & 1.00 & 0.24  \\
         \textbf{SEANet}       & 8.70M & \textbf{12.88} & \textbf{13.33} & \textbf{2.34} & \textbf{0.89} & \textbf{12.77} & \textbf{13.13} & \textbf{1.09} & \textbf{0.25} \\
         \hline
    \end{tabular}
    \label{Vox2}
\end{table*}

\subsubsection{\textbf{Comparison with other methods on the LRS2 or VoxCeleb2}}
Besides the three baseline methods, some other AV-TSE solutions are studied in previous works. Here, we compare SEANet with them from the corresponding published paper in Table~\ref{others}. We summarise the results in SI-SDR and SDR since most of these works focus on these two metrics. Note that the experimental settings and trial selections are slightly different in these works. So we remark the number of training data in Table~\ref{others}. Compared with the best previous works for the VoxCeleb2 dataset (12.13 dB from AV-Sepformer), SEANet uses fewer or equal samples for training and achieves 0.75 dB SI-SDR improvement. Similarly, for the LRS2 dataset, SEANet achieves 1.88 dB SI-SDR improvement compared to the previous method (11.3dB from Conversation). That guarantees our study's rigour. 

\begin{table}[!ht]
    \centering
        \caption{Comparison of SEANet and previous methods on the LRS2 dataset or VoxCeleb2 dataset. These results come from their original paper. $^{\dag}$ denotes that value is SDRi. `-' means that number is not reported in the original paper. }
    \begin{tabular}{ccccc}
        \hline
        Dataset & Method & SI-SDR & SDR & \# training data \\
        \hline
        \multirow{4}{*}{LRS2} & CaffNet-C~\cite{lee2021looking}   &       10.01 & - & - \\
        & LWTNet~\cite{afouras2020self}                & 10.80  & - & - \\
        & Conversation~\cite{afouras2018conversation}  & 11.30  & - & - \\
        & \textbf{SEANet} & {13.08}  & {13.67} & 20k \\
        \hline
        \multirow{8}{*}{VoxCeleb2} & Face filter~\cite{chung2020facefilter}   & - & 2.53$^{\dag}$ & 200k\\
        & Conversation~\cite{afouras2018conversation}  & - & 7.90 & - \\
        & VisualVoice~\cite{gao2021visualvoice}   & 9.73 & 10.20 & $>$ 1,000k \\
        & AV-Conv~\cite{wu2019time} & 10.38 & 10.90$^{\dag}$ & 40k \\
        & MuSE~\cite{pan2021muse, Lin2023sepformer} & 11.24  & - & 20k\\
        & AVDiffuSS~\cite{lee2023seeing}     & 12.03 & - & $>$ 1,000k \\
        & AV-Sep~\cite{Lin2023sepformer}  & 12.13  & - &  20k  \\
        & \textbf{SEANet} & {12.88} & {13.33} & 20k \\
        \hline
    \end{tabular}
    \label{others}
\end{table}

\subsection{\textbf{Cross-domain evaluation: LRS3, Grid and TCD-TIMIT}}
Then, we conduct the cross-domain evaluation for SEANet and the baseline methods on Table~\ref{cross} based on SI-SDR, SDR, PESQ and STOI. The evaluation sets are LRS3, Grid and TCD-TIMIT datasets. We find that the videos in the Grid dataset are not strictly synchronized. However, all videos in the LRS2 dataset are strictly synchronized. This setting mismatch can lead to a performance gap. Due to that, all methods are trained on the VoxCeleb2 for fair comparison. 

For the results on LRS3 and Grid, SEANet performs better than all baseline methods for all evaluation metrics. In the TCD-TIMIT dataset, AV-Sepformer performs slightly better than SEANet for two metrics. However, compared with AV-Sepformer, SEANet only used around 40\% number of model parameters. So extending the model size of SEANet can achieve better performance. In summary, the average SI-SDR and PESQ of SEANet for three datasets are over 13.81 dB and 2.42, which represents a good extraction quality. Considering the diversity of these evaluation sets, these experiments prove the robustness of our proposed SEANet and the value of exploring noise signals. 

\label{res-2}
\begin{table}[!htb]
    \centering
    \caption{Cross-domain evaluation of SEANet and the baseline methods, all methods are trained on VoxCeleb2, then test on LRS3, Grid and TCD-TIMIT datasets. Higher is better for all these metrics.}
    \begin{tabular}{p{1.3cm}<{\centering}p{1.4cm}<{\centering}p{1cm}<{\centering}p{.7cm}<{\centering}p{.7cm}<{\centering}p{.7cm}<{\centering}}
        \hline
        Evaluation & \multirow{2}{*}{{Method}} & \multicolumn{4}{c}{{Metrics}}\\
        Sets & & SI-SDR & SDR & PESQ & STOI \\
        \hline
        \multirow{4}{*}{LRS3} & AV-DPRNN  & 12.83 & 13.31 & 2.17 & 0.92 \\
                              & MuSE~\cite{pan2021muse}   & 12.58 & 13.08 & 2.15 & 0.92 \\
                              & AV-Sep~\cite{Lin2023sepformer}  & 14.52 & 14.96 & 2.40 & 0.93 \\
                              & \textbf{SEANet}  & {14.75} & {15.23} & {2.42} & {0.93} \\
        \hline
        \multirow{4}{*}{TCD} & AV-DPRNN  & 12.24 & 13.14 & 2.32 & 0.88 \\
                              & MuSE~\cite{pan2021muse}   & 12.57 & 13.25 & 2.33 & 0.89 \\
                              & AV-Sep~\cite{Lin2023sepformer}  & {14.15} & 14.81 & {2.49} & 0.91 \\
                              & \textbf{SEANet}  & {14.10} & {15.02} & {2.44} & {0.91} \\
         \hline
        \multirow{4}{*}{Grid} & AV-DPRNN  & 11.59 & 12.58 & 2.24 & 0.87 \\
                              & MuSE~\cite{pan2021muse}   & 11.19 & 12.19 & 2.27 & 0.88 \\
                              & AV-Sep~\cite{Lin2023sepformer} & 12.20 & 13.20 & 2.32 & 0.88 \\
                              & \textbf{SEANet}  & {12.59} & {14.45} & {2.43} & {0.88} \\
        \hline
    \end{tabular}
    \label{cross}
\end{table}

\subsection{\textbf{Ablation study for efficiency}}
\label{res-3}
This section conducts ablation studies to prove the efficiency of the proposed architecture and analyse why it can achieve better performance. We performed experiments on various architectures that were modified based on our framework or re-implemented from previous studies for comparison. Here, we consider AV-DPRNN for comparison since it shares the similar model architecture with our SEANet, the difference is that SEANet involves reverse attention to learn the speech-noise interaction. As we mentioned, some videos in VoxCeleb2 are out of sync by -0.08 to 0.08 seconds. Our ablation studies are conducted on the LRS2 dataset, since it is a high-quality audio-visual dataset with clean target speech.

\subsubsection{\textbf{Reverse attention}} 

\begin{figure}[!htb]
  \centering
  \includegraphics[width=\linewidth]{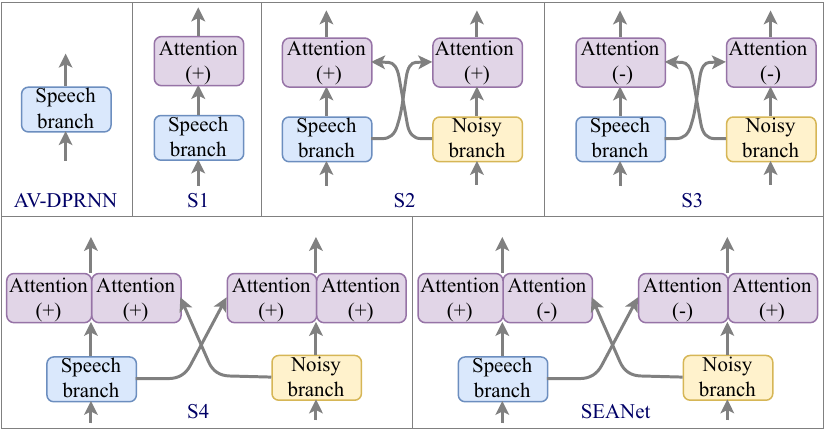}
  \caption{Four kinds of architecture for ablation study. Attention (+) denotes the normal self-attention, Attention (-) denotes the proposed reverse attention. S1 involves the self-attention for speech branch; S2 contains the positive cross-attention for speech-noise interaction; S3 contains the reverse attention for speech-noise interaction; similar to S2, S4 adds the self-attention in both speech branch and noisy branch.}
  \label{reverse}
\end{figure}

To verify that the proposed reverse attention is important to assist the extraction process, we consider the following structures in Figure~\ref{reverse} for comparison: \underline{AV-DPRNN} has the speech branch only. \underline{S1} contains the speech-speech interaction with attention (Eq.~\ref{Equation} contains self-attention $A_{s+}$ only). \underline{S2} has the noisy branch, the speech-noise interaction is conducted based on traditional self-attention (Eq.~\ref{Equation} contains $A_{s-}$ only without reverse operation). \underline{S3} is similar to S2, while speech-noise interaction based on reverse attention (Eq.~\ref{Equation} contains $A_{s-}$ only). \underline{S4} has speech-noise and speech-speech interaction attention, both of them apply traditional self-attention ($A_{s-}$ in Eq.~\ref{Equation} has no reverse operation). \underline{SEANet} contains the reverse attention for speech-noise interaction and self-attention for speech-speech interaction. 

We train these structures with the same setting and summarize the results in Table~\ref{reverse_results}: 1) Speech-speech attention can assist AV-TSE by learning voice consistency (AV-DPRNN vs S1). 2) The noisy branch can help the AV-TSE training by understanding the speech-noise interaction (AV-DPRNN vs S2 and S3). 3) Moreover, reverse attention performs slightly better than traditional self-attention in studying speech-noise interaction (S2 vs S3, S4 vs SEANet). 4) SEANet can achieve the best performance by incorporating reverse attention and self-attention.

We further involve the previous method as baseline for comparison. \cite{hu2022interactive, zheng2021interactive} also involve the noise during speech enhancement with a straightforward CNN-based interaction solution. We re-implement their approach in our AV-TSE system to achieve a fair comparison. As shown in Table~\ref{reverse_results}, SEANet performs better than~\cite{hu2022interactive, zheng2021interactive} since it explores more potential of speech-noise interaction.

\begin{table}[!ht]
    \centering
        \caption{Ablation study for the different network structures. $\oplus$ denote typical self-attention, which is positive; $\ominus$ represents our proposed reverse attention, which is negative. Experiments are conducted on the LRS2 dataset. We also re-implement the previous method in AV-TSE as the baseline for comparision. Illustrations of these structures can be found in Figure~\ref{reverse}}
    \begin{tabular}{ccccc}
    \hline
        \multirow{2}{*}{Method} & Speech-noise & Speech-speech & \multirow{2}{*}{SI-SDR} & \multirow{2}{*}{SDR} \\
        & attention  & attention   &  & \\
        \hline
{\cite{hu2022interactive, zheng2021interactive}} & {NA} & {NA} & {11.10} & {11.67} \\
        AV-DPRNN & \xmark & \xmark & 10.24 & 10.81\\
        S1 & \xmark & $\oplus$ & 12.18 & 12.79\\
        S2 & $\oplus$ & \xmark & 12.51 & 13.05\\
        S3 & $\ominus$ & \xmark & 12.71 & 13.33\\
        S4 & $\oplus$ & $\oplus$ & 12.75 & 13.39\\
        \textbf{SEANet} & $\ominus$ & $\oplus$ & {13.08} & {13.67} \\
        \hline
    \end{tabular}
    \label{reverse_results}
\end{table}

\subsubsection{\textbf{Model size}} 

As we mentioned, SEANet can perform better than MuSE and AV-Sepformer with the smaller model size: 8.70M (SEANet) vs 15.01M (MuSE) and 21.73M (AV-Sepformer). Considering that SEANet has more number parameters than the AV-DPRNN, which shares a similar RNN-based architecture except involving the noisy branch, Table~\ref{abl-size} propose two new structures to study the efficiency of SEANet further. `-deeper' means AV-DPRNN with more layers ($R$) and `-wider' represents AV-DPRNN with the larger feature dimension ($D$). These two systems have more parameters than SEANet, while from the results on LRS2, SEANet still achieves better SI-SDR and SDR with over 1.5 dB difference. These experiments prove that our improvement is not parameters-driven.

\begin{table}[!ht]
    \centering
        \caption{Ablation study for the model size, SEANet can perform better than the deeper or larger AV-DPRNN with the smaller architecture.}
    \begin{tabular}{ccccc}
    \hline
        Method & \# Param & SI-SDR & SDR & Setting \\
        \hline
        AV-DPRNN  & 4.12M & 10.24 & 10.81 & $R = 5,  D = 64 $ \\
        -deeper   & 9.32M & 11.25 & 11.80 & $R = 14, D = 64 $ \\
        -wider    & 11.88M & 11.45 & 12.01 & $R = 5,  D = 128 $ \\
        \hline
        \textbf{SEANet}    & 8.70M & {13.08} & {13.67} & $R = 5, D = 64 $ \\
        \hline
    \end{tabular}
    \label{abl-size}
\end{table}

\subsubsection{\textbf{Extraction correctness}} 
\begin{figure}[!ht]
  \centering
  \includegraphics[width=.95\linewidth]{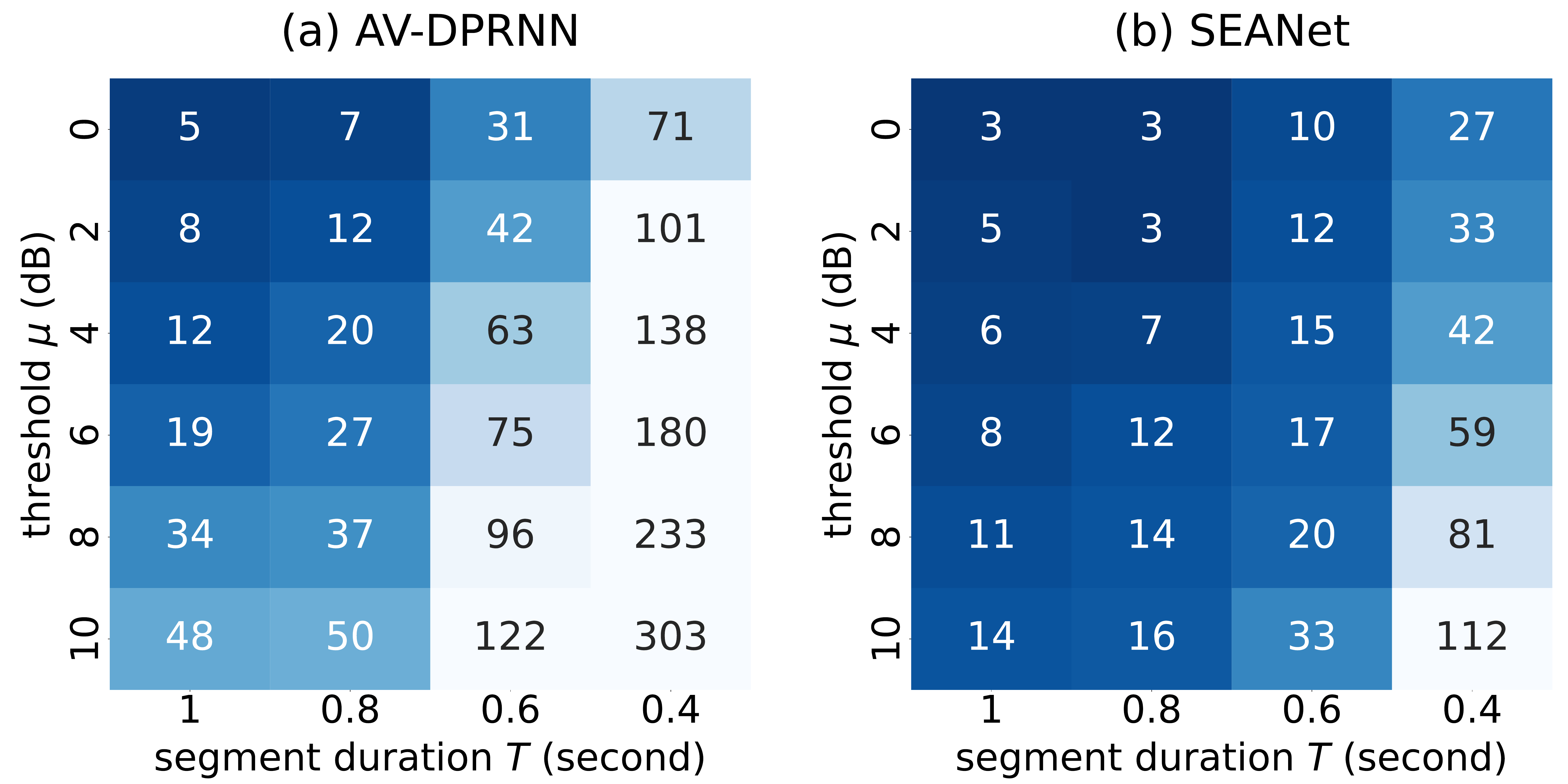}
  \caption{The total number of INCORRECT speaker extraction segments during evaluation. Larger number represents more incorrect extractions.}
  \label{Correctness_study}
\end{figure}

Then, we study whether SEANet obtains the improvement by achieving more correct extraction. During the evaluation, we split the speech into segments of fixed duration $T$. Consider $l(\hat{s}, s)$ as the SI-SDR between the estimated speech $\hat{s}$ and the ground-truth speech $s$. We define that the extraction in one audio segment is incorrect if the extracted audio is much close to the ground-truth noise $n$ instead of the ground-truth speech $s$: $ l(\hat{s}, s) < l(\hat{s}, n) - \mu$ Here, we involve $\mu$ as the threshold where Lower $T$ and higher $\mu$ will lead to more stringent requirements and incorrect extraction segments. Figure~\ref{Correctness_study} shows the total number of incorrect extraction segments during the evaluation. Compared with the results of AV-DPRNN on the left side, SEANet leads to less incorrect extraction for all the settings. That verifies our assumption that SEANet can alleviate incorrect speaker extraction to improve performance.

\begin{figure*}[!htb]
  \centering
    \includegraphics[width=\linewidth]{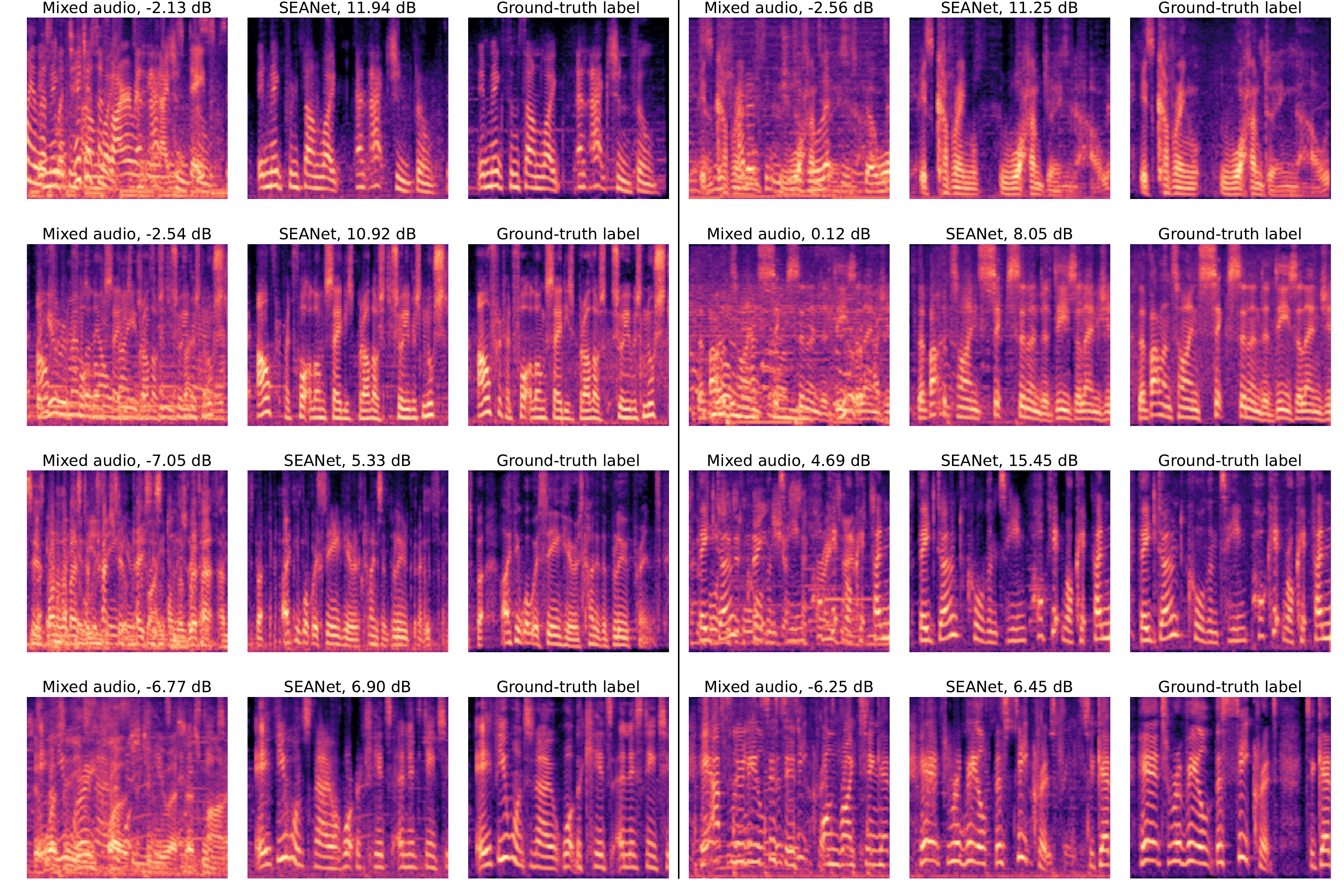}
  \caption{SEANet extraction visualization for 8 samples in the LRS2 with SI-SDR. Each figure represents the Mel spectrogram of the mixed audio and the SEANet's output speech. The corresponding SI-SDR is also provided. Compared with the mixed audio, SEANet is able to filter the noise and improve the SI-SDR.}
  \label{Vismore}
\end{figure*}

\subsubsection{\textbf{Model architecture}}
We further conducted experiments for SEANet with three different model architectures. As shown in Table~\ref{r11}, SEANet$^\alpha$ performs much worse than SEANet due to the lack of extractor and suppressor (13.08 dB VS 10.57 dB). With pre-extractor and pre-suppressor only , SEANet can hardly learn high-quality speech representation. Meanwhile, SEANet$^\beta$ performs slightly worse than SEANet, which means the noise-noise self-attention is useful in SEANet to guide the system to learn the noisy representation (13.08 dB VS 12.69 dB). That can further boost the speech representation through the speech-noise interaction block. SEANet$^\gamma$ modifies the order of subtraction and softmax in Eq~\ref{Equation}: 
\begin{equation}
    A_{s} = softmax{(\frac{(Q_s-Q^{\prime}_n) \cdot {K_s}^T}{\sqrt{D}}})
    \label{Equation2}
\end{equation}
It achieves the similar result with SEANet (13.08 dB VS 12.98 dB). That verifies both implementation approaches are useful here. These experiments prove the reasonableness of SEANet.

\begin{table}[!ht]
    \centering
        \caption{Ablation study for the PSNL block on the LRS2 dataset. Compared with SEANet, SEANet$^\alpha$ denotes to remove the extractor and suppressor in PSNL block, SEANet$^\beta$ denotes to remove the noise-noise self-attention in the suppressor, SEANet$^\gamma$ denotes to do subtraction on the logits first then feed into softmax.}
        \begin{tabular}{cccc}
        \hline
        Method & SI-SDR & SDR & Remark\\
        \hline
        \textbf{SEANet}   & 13.08 & 13.67 & -\\
        \textbf{SEANet$^\alpha$}  & 10.57 & 11.17 & Remove extractor/suppressor\\
        \textbf{SEANet$^\beta$}   & 12.69 & 13.31 & Remove noise-noise self-attention\\
        \textbf{SEANet$^\gamma$}   & 12.98 & 13.54 & Do subtraction before softmax \\
    \hline
    \end{tabular}
    \label{r11}
\end{table}

\subsubsection{\textbf{Model efficiency}}
We further study the complexity of the SEANet model and compare it with two popular structures: CNN-based MuSE~\cite{pan2021muse} and Transformer-based AV-Sepformer~\cite{Lin2023sepformer}. As shown in Table~\ref{abl-com}. Our SEANet achieves the best performance with the relative small number of parameters and lower FLOPs. Furthermore, we conduct the experiments to study the relationship among model size, complexity and the performance in SEANet. SEANet$^s$ and SEANet$^l$ denote the small version SEANet and large version SEANet, respectively.

\begin{table}[!ht]
    \centering
        \caption{Ablation study for the model complexity on the LRS2 dataset, $R$ and $D$ denote the depth of layers and dimension of layers in the SEANet, respectively.}
    \begin{tabular}{ccccc}
    \hline
        Method & \# Param & FLOPs & SI-SDR & Setting \\
        \hline
        MuSE~\cite{pan2021muse}          & 15.01M & 34.23G & 10.97 & - \\
        AV-Sepformer~\cite{Lin2023sepformer}  & 21.73M & 73.09G & 12.60 & - \\
        \hline
        \textbf{SEANet$^{D-}$}  & 3.32M & 8.46G & 10.71 & $R = 5, D = 32$ \\
        \textbf{SEANet$^{R-}$}  & 5.12M & 9.75G & 11.25 & $R = 2, D = 64$ \\
        \textbf{SEANet}         & 8.70M & 21.17G & 13.08 & $R = 5, D = 64$ \\
        \textbf{SEANet$^{R+}$}  & 12.28M & 36.59G & 13.69 & $R = 8, D = 64$ \\
        \textbf{SEANet$^{D+}$}  & 30.10M & 72.72G & 14.68 & $R = 5, D = 128$ \\
        \hline
    \end{tabular}
    \label{abl-com}
\end{table}

\subsection{\textbf{Quality study for robustness}}
\label{res-4}

In this section, we conduct quality studies to learn the robustness of the proposed SEANet under the different training or evaluation settings. Similarly, we evaluate the AV-DPRNN system for comparison on the LRS2 dataset since it contains the clean target speech.

\subsubsection{\textbf{Hyper-parameter}} 
\begin{figure}[!ht]
  \centering  
  \includegraphics[width=\linewidth]{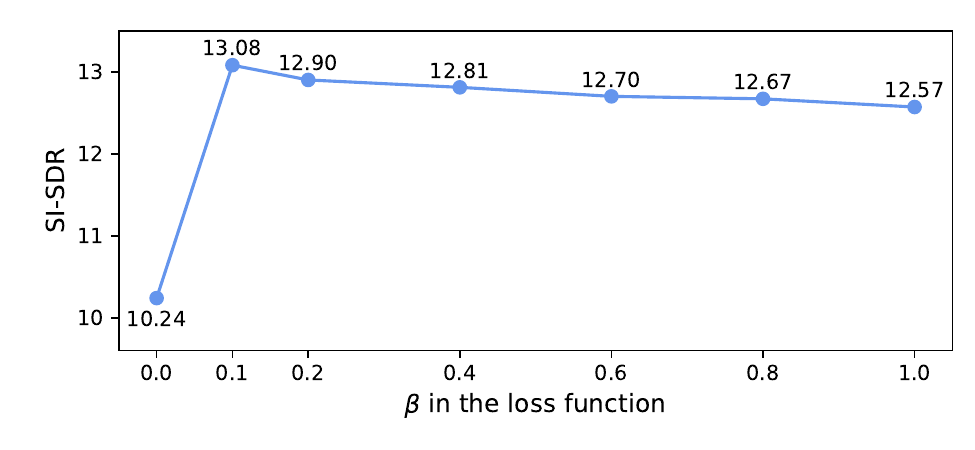}  
  \caption{Study for the effect of hyper-parameter $\beta$ in the loss function for SEANet. SEANet degrades to baseline AV-DPRNN when $\beta = 0$.}
  \label{Hyper_parameter_study}
\end{figure}

Figure~\ref{Hyper_parameter_study} studies the robustness of SEANet when training with the different hyper-parameter $\beta$ in the loss function Eq.~\ref{Equation2}. Note that SEANet degrades to AV-DPRNN when $\beta$ is equal to zero. From this figure, SEANet is stable under the different $\beta$ over zero. A lower value of $\beta$ can lead to better performance, and SEANet can achieve the best performance when $\beta$ equals 0.1. That proves our SEANet is not sensitive to the hyper-parameter selection.

\subsubsection{\textbf{AV-TSE for ordinary acoustic environment}}
Then, we report the results of SEANet for ordinary acoustic environment: when the mixture of audio contains the speech from the target speaker and the non-speech noise only, our task can be viewed as audio-visual speech enhancement. We train the AV-DPRNN and SEANet based on these speech enhancement format data for comparison. The non-speech noisy signals are sampled from the MUSAN dataset~\cite{MUSAN}, which contains the nature noises (the sounds from thunder, rain, ball, etc) and the background music (playing the instrument or singing). The SNR is randomly selected from -5 dB to 5 dB. In Table~\ref{tab:audio-only}, we report the results with the different Signal-to-Noise Ratio (SNR) for the input audio. SEANet performs better than AV-DPRNN for this enhancement task under different noise levels with the very stable improvement. Meanwhile, a very high SI-SDR can be achieved in this enhancement task. Because extracting the target speaker's speech from the non-speech noise is easier than extracting it from the mixture of speech.

\begin{table}[!ht]
    \centering
    \vspace{-2mm}   
    \caption{Performance of AV-DPRNN and the proposed SEANet in the ordinary acoustic environment for audio-visual speech enhancement task. The input audio contains the target speaker's speech and the non-speech noise. Input SNR (dB) indicates the SNR for the input audio to represent the different noisy level.}
    \begin{tabular}{cccccc}
    \hline
        Method & Input SNR (dB) & SI-SDR & SDR & SI-SDRi & SDRi \\  
        \hline
        \multirow{4}{*}{AV-DPRNN} & 5  & 13.86 & 14.42 & 8.75 & 9.23\\
                                  & 10 & 16.96 & 17.51 & 6.85 & 7.34\\
                                  & 15 & 19.68 & 20.31 & 4.57 & 5.14\\
                                  & 20 & 21.77 & 22.57 & 1.66 & 2.40\\
        \hline
        \multirow{4}{*}{\textbf{SEANet}}   & 5  & 14.67 & 15.23 & 9.57 & 10.04\\
                                  & 10 & 17.67 & 18.21 & 7.57 & 8.03\\
                                  & 15 & 20.39 & 20.99 & 5.29 & 5.82\\
                                  & 20 & 22.36 & 23.06 & 2.25 & 2.88\\
        \hline
    \end{tabular}
    \label{tab:audio-only}
\end{table}

\subsubsection{\textbf{AV-TSE for diverse acoustic environments}}
Our previous results have shown that SEANet can handle the scenario of two speakers talking simultaneously. Here, we further study the robustness of our SEANet in diverse acoustic environments on the LRS2 dataset. As shown in Table~\ref{tab:aug}, the audio mixture contains the target speaker's speech. Then, `S' denotes the existence of one interference speaker. `N' means the existence of non-speech noise, such as background noise or music. For instance, `S + S + N' represents the input audio containing the target speaker's voice, two other speakers' voices and the non-speech noise. This can simulate very complex extraction conditions.

In this experiment, instead of training with the mixed audio of two speakers' voices only, both AV-DPRNN and SEANet are trained with the data from these four scenarios to boost the application ability. From the results, a more noisy environment leads to a low SI-SDR and SDR since the extraction becomes challenging. At the same time, the SI-SDRi and SDRi are relatively stable. Compared with AV-DPRNN, SEANet shows more excellent stability under all conditions. In these challenge scenarios, it consistently achieves impressive SISDR, SDR, SISDRi, and SDRi. 

\begin{table}[!ht]
    \centering    
    \caption{Performance of AV-DPRNN and the proposed SEANet in the challenging acoustic environment. For the mixture of audio which contains the target speaker's speech, `S' denotes the existence of one interference speaker, and `N' denotes the existence of the non-speech noisy signal.}
    \begin{tabular}{p{1.4cm}<{\centering}p{1.5cm}<{\centering}p{.9cm}<{\centering}p{.7cm}<{\centering}p{1cm}<{\centering}p{.7cm}<{\centering}}
    \hline
        Method & Interference & SI-SDR & SDR & SI-SDRi & SDRi \\
        \hline
        \multirow{4}{*}{AV-DPRNN} & S & 10.44 & 11.13 & 10.49 & 10.97\\
                                  & S + N & 8.93 & 9.64 & 9.86 & 10.36\\
                                  & S + S & 5.48 & 6.35 & 9.95 & 10.50\\
                                  & S + S + N & 4.93 & 5.82 & 9.71 & 10.27\\
        \hline
        \multirow{4}{*}{\textbf{SEANet}}   & S  & 13.21 & 14.05 & 13.26 & 13.89\\
                                  & S + N  & 11.37 & 12.21 & 12.31 & 12.94\\
                                  & S + S & 7.94 & 8.89 & 12.41 & 13.04\\
                                  & S + S + N & 7.20 & 8.16 & 11.98 & 12.60\\
        \hline
    \end{tabular}
    \label{tab:aug}
\end{table}

\subsection{\textbf{Multi-modal interaction}}

In this section, we provide the results of the proposed multi-modal interaction method on the LRS2 dataset. All experiments are fine-tuned on the pre-trained SEANet structure to speed up the training process.

As shown in Table~\ref{SEANet+}, all three methods get robust performance for SI-SDR and SDR. The best one is F-SEANet, which achieves 13.63 dB SI-SDR, P-SEANet and A-SEANet achieve 13.36 dB SI-SDR and 13.51 dB SI-SDR, respectively. These results show that involving vision information in the early stage can benefit the system more, while the late stage of involvement can hardly boost the system.

We propose the new assumptions to the importance of visual information in AV-TSE. In the long-term research process, one consensus is that lip movements can guide the extraction system in understanding the detailed target speech representation along the time axis. However, the lip movements may play the role of the `mentor' to hint the system knows which kind of voice should be extracted with a global view. The evidence is that it is more important to involve lip movements in the early stage of the extraction network than in the late stage. In the future, we will focus more on the efficient design in the early-stage audio-visual fusion to confirm whether our arguments are correct.

\begin{table}[!ht]
    \centering
        \caption{Ablation study for the audio-visual interaction learning based on SEANet. The results are conducted on the LRS2 dataset.}
    \begin{tabular}{cccc}
    \hline
        \multirow{1}{*}{Method} & \multirow{1}{*}{SI-SDR} & \multirow{1}{*}{SDR} & Remark \\
        \hline
        AV-DPRNN & 10.24 & 10.81 & -\\
        MuSE & 10.97 & 11.57 & -\\
        AV-Sep & 12.60 & 13.13 & - \\
        \textbf{SEANet} & 13.08 & 13.67 & - \\
        \hline
        \textbf{F-SEANet} & 13.63 & 14.20 & Fusion-based \\
        \textbf{P-SEANet} & 13.36 & 13.93 & PSNL-based  \\
        \textbf{A-SEANet} & 13.51 & 14.08 & Algorithm-based\\
        
        \hline
    \end{tabular}
    \label{SEANet+}
\end{table}

\subsection{\textbf{Visualization}}
\label{res-5}
In this section, we provide some visualization results of SEANet. These results are also achieved on the LRS2 dataset.

Figure~\ref{Vismore} visualizes the extraction results of eight random audio samples. Each line contains two samples. For each sample, we provide the Mel spectrogram of the mixed audio, the SEANet's output speech and the ground truth speech for comparison. The SI-SDR for the mixed audio and the output speech has also been provided. Compared with the mixed audio, the output speech resembles the ground-truth speech. These visualization results demonstrate that SEANet can achieve robust extraction in both easy (high original SI-SDR) and challenging (low original SI-SDR) scenarios.



\section{Discussion}
\subsection{Extraction vs separation}
For audio-visual target speaker extraction (AV-TSE)~\cite{pan2021muse, afouras2018conversation} and audio-visual speech separation~\cite{ephrat2018looking}. Although both require information from face frames, 1) the separation model needs the face from multiple sound sources. In contrast, AV-TSE requires the target face only, which is a more convenient solution. 2) during training and testing, the separation model requires the same number of speakers due to the fixed model structure, while the extraction model can handle any number of speakers by applying the model multiple times. In summary, AV-TSE is a more feasible direction for diverse real-world conversations. 

\subsection{Video quality for AV-TSE}
We find that the audio in VoxCeleb2 is not clean enough because it is collected from the interviewers with background music, noise, or even other speakers' speech. Therefore, using such speech as the flawed ground-truth label will damage many extraction systems since we usually assume the ground-truth label is absolutely correct. Furthermore, VoxCeleb2 has many out-of-sync videos, while AV-TSE models are designed based on the strictly synchronized condition. Previous study~\cite{lee2021looking} has verified that the audio-visual offset will affect AV-TSE. It is important to analyze the advantages and disadvantages of using such an imperfect dataset for AV-TSE. Meanwhile, studying the extraction under challenging real-world conditions, such as discontinuous face frames and the boisterous environment, is also meaningful.

\section{Conclusion}
In this paper, we propose the Subtraction-and-ExtrAction network (SEANet) for audio-visual target speaker extraction (AV-TSE). SEANet focuses on selective auditory attention to force the neural network to understand the noisy signal during extraction. Our experimental results verify that this `subtraction' strategy can suppress noisy signal and avoid incorrect speaker extraction. SEANet outperforms previous methods for five different datasets under up to nine metrics. In future work, we will extend the selective auditory attention mechanism to the audio-only target speaker extraction. Furthermore, we will analyse the roles of each auxiliary reference to achieve an intelligent balance between them.

\ifCLASSOPTIONcaptionsoff
  \newpage
\fi

\bibliographystyle{IEEEtran}
\bibliography{IEEEabrv,Bibliography}

\begin{IEEEbiography}[{\includegraphics[width=1in,height=1.25in]{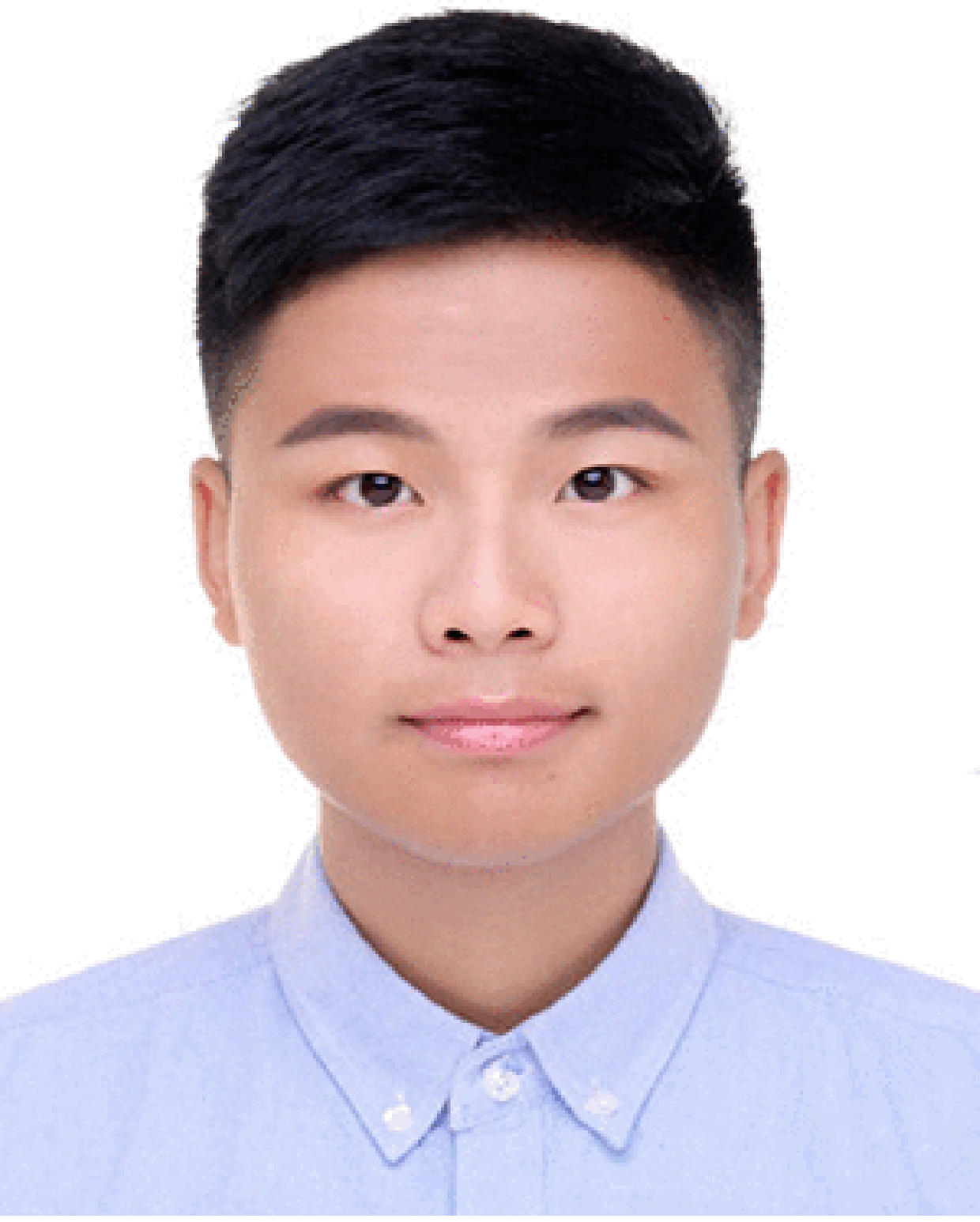}}]{Ruijie Tao} (Member, IEEE) received the Ph.D. and M.Sc. degree from National University of Singapore, Singapore, in 2023 and 2019, respectively. He received the B.Eng. degree from Soochow University, China, in 2018. He is currently a research fellow at National University of Singapore, Singapore. He is also the reviewer of CVPR, ICASSP, Interspeech, SPL, CSL and SLT. His research interests include audio-only and audio-visual speaker recognition, active speaker detection, speaker diarization, speech enhancement, speech extraction and anti-spoofing.
\end{IEEEbiography}

\begin{IEEEbiography}[{\includegraphics[width=1in,height=1.25in,clip,keepaspectratio]{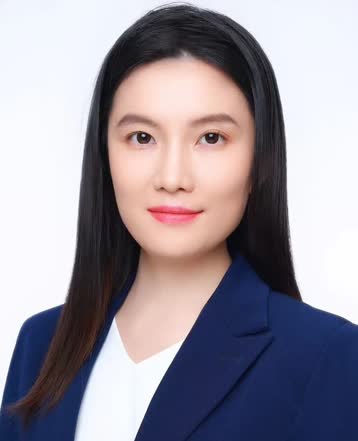}}]{Xinyuan Qian} (Senior Member, IEEE) received the B.Eng (First class) and M.Sc. (Distinction) degrees from University of Edinburgh, and the Ph.D. degree in Computer Science from Queen Mary University of London, U.K. During the Ph.D. study, she has been the visiting international student in FBK, trento, Italy. She has then worked as a research fellow at National University of Singapore, Singapore and The Chinese University of Hong Kong, Shenzhen. Since 2022, she joined University of Science and Technology Beijing, China as the Associate Professor. Her research interests include audio-visual fusion, speaker localization and tracking, speaker extraction, speech synthesis and recognition. 
\end{IEEEbiography}

\begin{IEEEbiography}[{\includegraphics[width=1in,height=1.25in]{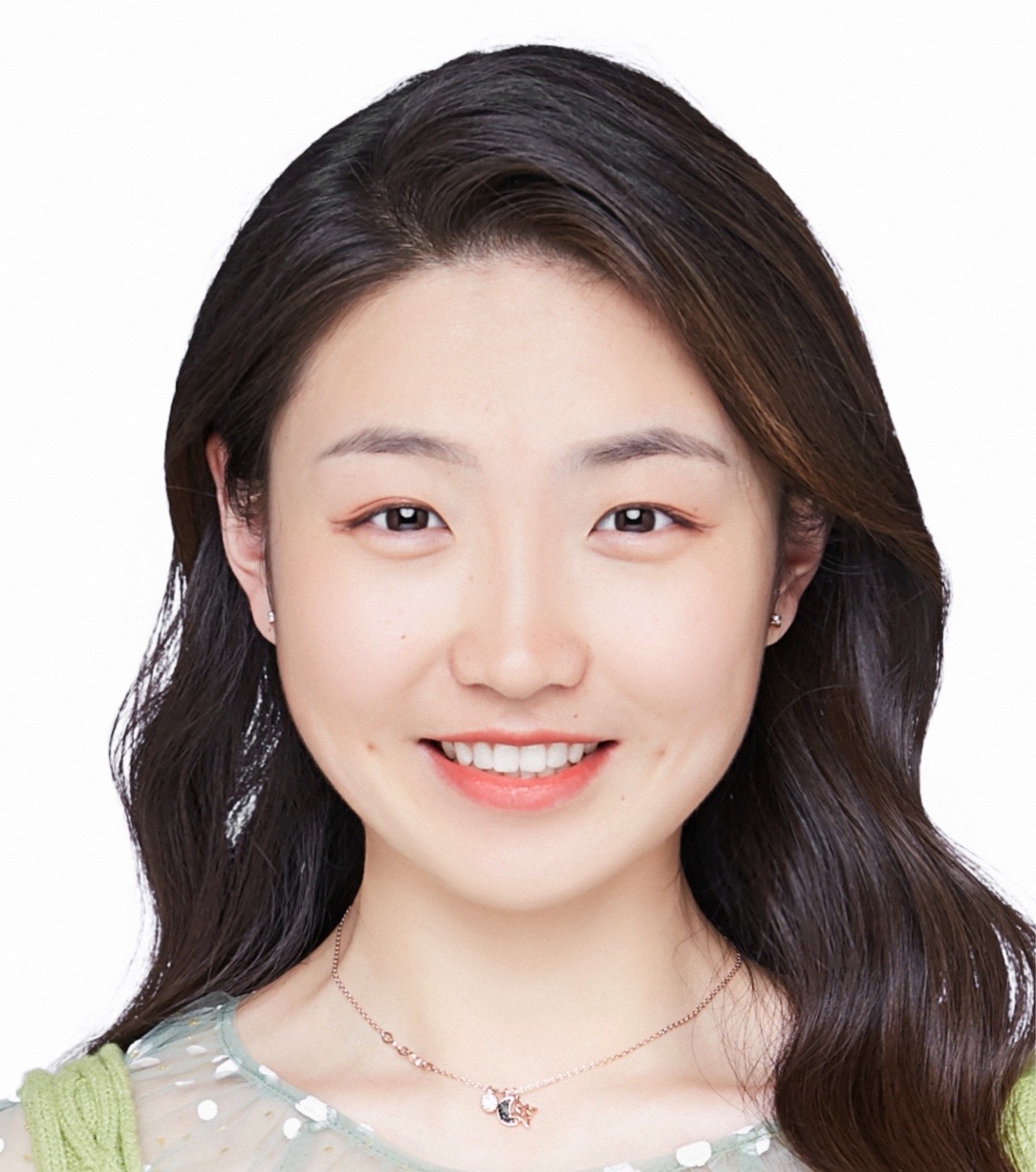}}]{Yidi Jiang} (Member, IEEE) 
received the M.Sc. from National University of Singapore, Singapore, in 2021, and the B.Eng. degree from University of Electronic Science and Technology of China, China, in 2020. She is currently a Ph.D. candidate at National University of Singapore, Singapore. She is also reviewer of ACM MM, ICASSP, Interspeech, IJCNN. Her research interests include active speaker detection, speaker diarization, speaker verification, dataset distillation.
\end{IEEEbiography}

\begin{IEEEbiography}[{\includegraphics[width=1in,height=1.25in]{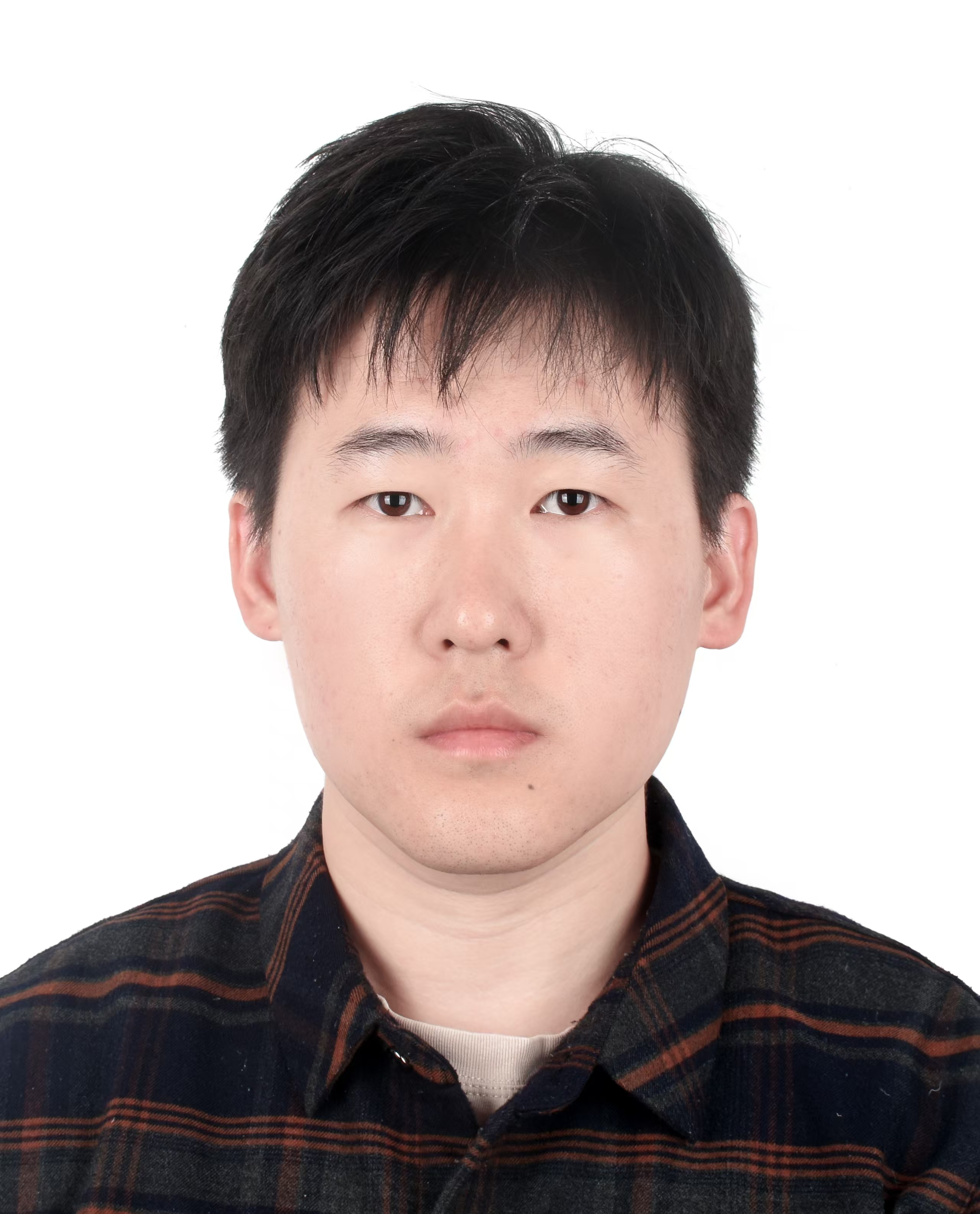}}]{Junjie Li} received the B.Eng. degree in Computer Science and Technology from Tiangong University, Tianjin, China, in 2020, and the M.Eng. degree in Electronics and Information from Tianjin University, Tianjin, China, in 2023. He is currently a Ph.D. candidate at Department of Electrical and Eletronic Engineering, Faculty of Engineering, The Hong Kong Polytechnic University, Hong Kong, China. His research interests include source separation, speech extraction,
speech enhancement, automatic speaker verification. 
\end{IEEEbiography}

\begin{IEEEbiography}[{\includegraphics[width=1in,height=1.25in,clip,keepaspectratio]{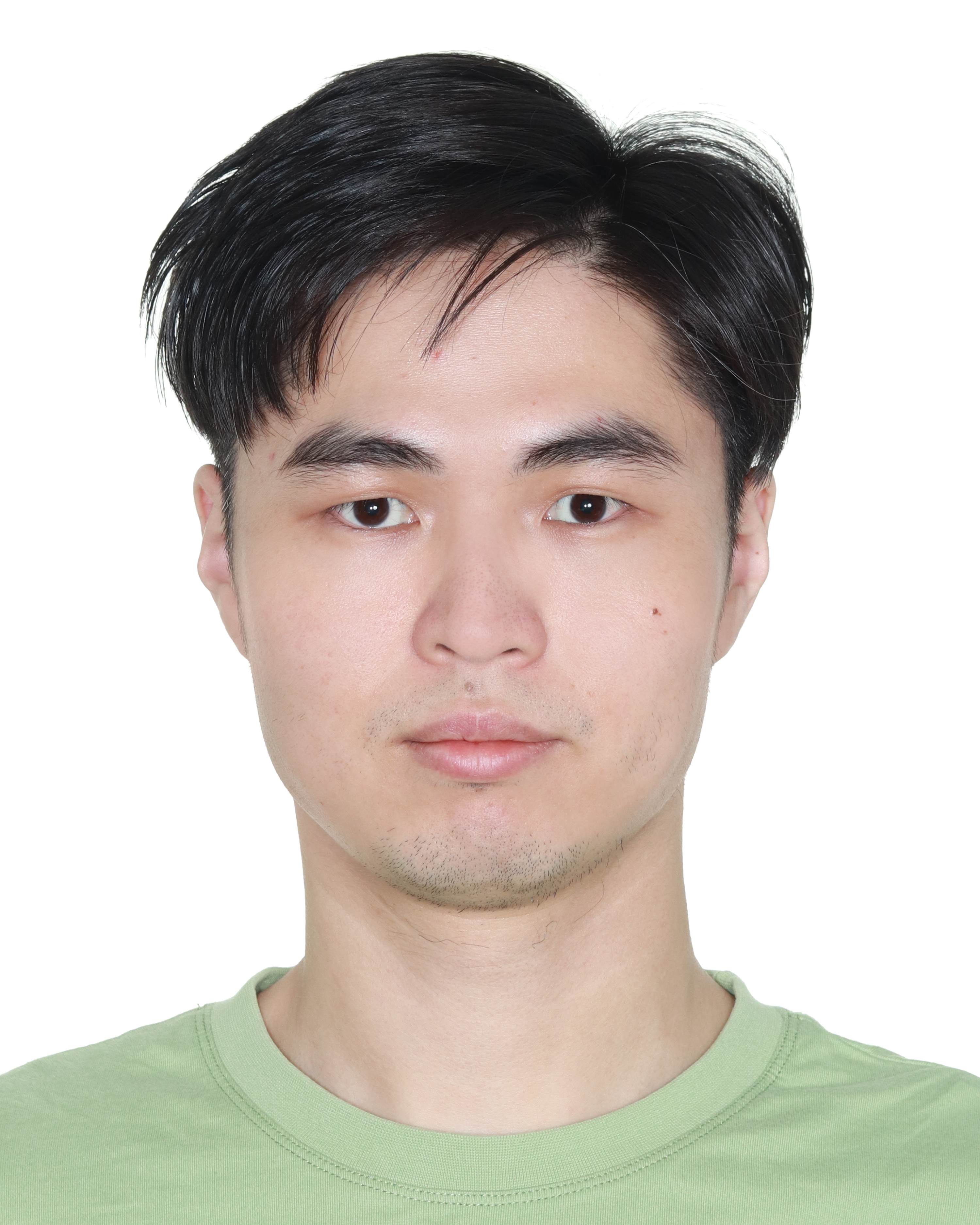}}]{Jiadong Wang} (Member, IEEE) received the Ph.D. degree from National University of Singapore, Singapore, in 2024, and the master’s degree from Zhejiang University, China, in 2019. and the bachelor’s degree from Northeastern University, China, in 2016. He is currently a research fellow at MRI, Technical University of Munich, Germany. Prior to that, he was a research assistant in Chinese University of Hong kong, Shenzhen, China. He is also a reviewer of ICCV, AAAI, ACM MM, ECCV, IROS. His research interests include audio-visual speech recognition, speaker extraction, sound source localization and tracking, talking face generation, spiking neural network.
\end{IEEEbiography}

\begin{IEEEbiography}[{\includegraphics[width=1in,height=1.25in]{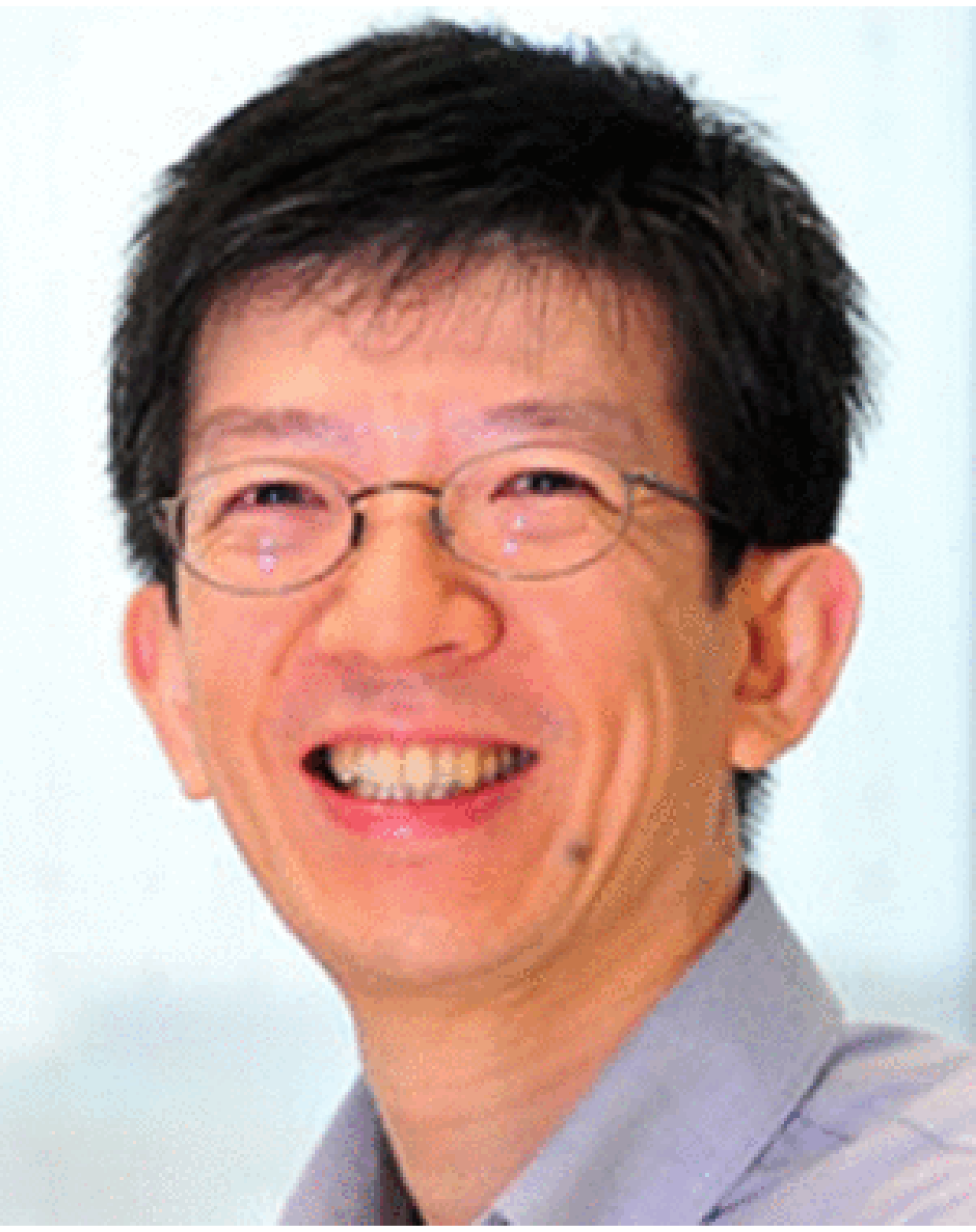}}]{Haizhou Li} (Fellow, IEEE) received the B.Sc., M.Sc., and Ph.D. degrees in electrical and electronic engineering from the South China University of Technology, Guangzhou, China, in 1984, 1987, and 1990 respectively. He is currently a Presidential Chair Professor and the Executive Dean of the School of Data Science, The Chinese University of Hong Kong, Shenzhen, China. He is also an Adjunct Professor with the Department of Electrical and Computer Engineering, National University of Singapore, Singapore. Prior to that, he taught with the University of Hong Kong, Hong Kong, (1988–1990) and South China University of Technology, (1990–1994). He was a Visiting Professor with CRIN in France (1994–1995), Research Manager with the AppleISS Research Centre (1996–1998), Research Director with Lernout \& Hauspie Asia Pacific (1999–2001), a Vice President with InfoTalk Corp. Ltd. (2001–2003), and the Principal Scientist and Department Head of Human Language Technology with the Institute for Infocomm Research, Singapore (2003–2016). His research interests include automatic speech recognition, speaker and language recognition, natural language processing. Dr. Li was an Editor-in-Chief of IEEE/ACM Transactions on Audio, Speech and Language Processing (2015–2018), a Member of the Editorial Board of Computer Speech and Language since 2012, an elected Member of IEEE Speech and Language Processing Technical Committee (2013–2015), the President of the International Speech Communication Association (2015–2017), the President of Asia Pacific Signal and Information Processing Association (2015–2016), and the President of Asian Federation of Natural Language Processing (2017–2018). He was the General Chair of ACL 2012, INTERSPEECH 2014, ASRU 2019 and ICASSP 2022. Dr. Li is a Fellow of the ISCA, and a Fellow of the Academy of Engineering Singapore. He was the recipient of the National Infocomm Award 2002, and the President's Technology Award 2013 in Singapore. He was named one of the two Nokia Visiting Professors in 2009 by the Nokia Foundation, and U Bremen Excellence Chair Professor in 2019.
\end{IEEEbiography}

\vfill

\newpage

\end{document}